\documentclass[sigconf, nonacm]{acmart}
\usepackage{algorithm}
\usepackage{algpseudocode}
\usepackage{amsmath}
 \usepackage{multirow}
\usepackage{enumerate}
\usepackage{graphicx}
 \usepackage{enumitem} 
 \usepackage{tabularx}
 \usepackage{booktabs} 
 \usepackage{diagbox}
 \usepackage{booktabs,siunitx,caption,dsfont}
 \usepackage[acronym,nohypertypes={acronym}]{glossaries}
\newacronym{rl}{RL}{\textit{Reinforcement Learning}}
\newacronym{sla}{SLA}{\textit{Service Level Agreement}}
\newacronym{tps}{TPS}{\textit{Transactions Processed per Second}}
\newacronym{qps}{QPS}{\textit{Queries Processed per second}}
\newacronym{dbms}{DBMS}{\textit{Database Management Systems}}
\newacronym{dba}{DBA}{\textit{Database Administrator}}
\newacronym{vm}{VM}{\textit{Virtual Machine}}
\newacronym{ml}{ML}{\textit{Machine Learning}}
\newacronym{dnn}{DNN}{\textit{Deep Neural Network}}
\newacronym{llm}{LLM}{\textit{Large Language Model}}
\newacronym{pca}{PCA}{\textit{Principal Component Analysis}}
\newacronym{ddpg}{DDPG}{\textit{Deep Deterministic Policy Gradients}}
\newacronym{drl}{DRL}{\textit{Deep Reinforcement Learning}}
\newacronym{tco}{TCO}{\textit{Total Cost of Ownership}}
\newacronym{ppo}{PPO}{\textit{Proximal Policy Optimization}}
\newacronym{rfecv}{RFECV}{\textit{Recursive Feature Elimination with Cross-Validation}}
\newacronym{lrt}{LRT}{\textit{Likelihood Ratio Test}}
\newacronym{bo}{BO}{\textit{Bayesian Optimization}}
\newacronym{CART}{CART}{\textit{Classification And Regression Tree}}

\newcommand\vldbdoi{10.14778/3785297.3785302}
\newcommand\vldbpages{589 - 602}
\newcommand\vldbvolume{19}
\newcommand\vldbissue{4}
\newcommand\vldbyear{2025}
\newcommand\vldbauthors{\authors}
\newcommand\vldbtitle{\shorttitle} 
\newcommand\vldbavailabilityurl{https://github.com/Orange-OpenSource/dot}
\newcommand\vldbpagestyle{empty} 

\begin{document}

\title{DOT: Dynamic~Knob~Selection and Online~Sampling for~Automated~Database~Tuning}
\author{Yifan Wang}
\affiliation{%
  \institution{Orange / Inria / Univ.\,Lille, France}
   \country{France}
}
\email{yifan.wang.etu@univ-lille.fr}

\author{Debabrota Basu}
\affiliation{%
  \institution{Inria / Univ.\,Lille / CNRS, France}
   \country{France}
}\email{debabrota.basu@inria.fr}

\author{Pierre Bourhis}
\affiliation{%
  \institution{CNRS / Univ.\,Lille / Inria, France}
   \country{France}
}\email{pierre.bourhis@univ-lille.fr}

\author{Romain Rouvoy}
\affiliation{%
  \institution{Univ.\,Lille / CNRS / Inria, France}
  \country{France}
}
\email{romain.rouvoy@univ-lille.fr}

\author{Patrick Royer}
\affiliation{%
  \institution{Orange, France}
   \country{France}
}
\email{patrick.royer@orange.com}

\begin{abstract}

\gls{dbms} are crucial for efficient data management and access control, but their administration remains challenging for \glspl{dba}.
Tuning, in particular, is known to be difficult. 
Modern systems have many tuning parameters, but only a subset significantly impacts performance. 
Focusing on these influential parameters reduces the search space and optimizes performance. 
Current methods rely on costly warm-up phases and human expertise to identify important tuning parameters.
In this paper, we present \textsf{DOT}, \emph{a dynamic knob selection and online sampling \gls{dbms} tuning algorithm}.
\textsf{DOT} uses \gls{rfecv} to prune low-importance tuning parameters and a \gls{lrt} strategy to balance exploration and exploitation.
For parameter search, \textsf{DOT} uses a \gls{bo} algorithm to optimize configurations on-the-fly, eliminating the need for warm-up phases or prior knowledge (although existing knowledge can be incorporated).
Experiments show that \textsf{DOT} achieves matching or outperforming performance compared to state-of-the-art tuners while substantially reducing tuning overhead.

\end{abstract}

\maketitle

\pagestyle{\vldbpagestyle}
\begingroup\small\noindent\raggedright\textbf{PVLDB Reference Format:}\\
\vldbauthors. \vldbtitle. PVLDB, \vldbvolume(\vldbissue): \vldbpages, \vldbyear.\\
\href{https://doi.org/\vldbdoi}{doi:\vldbdoi}
\endgroup
\begingroup
\renewcommand\thefootnote{}\footnote{\noindent
This work is licensed under the Creative Commons BY-NC-ND 4.0 International License. Visit \url{https://creativecommons.org/licenses/by-nc-nd/4.0/} to view a copy of this license. For any use beyond those covered by this license, obtain permission by emailing \href{mailto:info@vldb.org}{info@vldb.org}. Copyright is held by the owner/author(s). Publication rights licensed to the VLDB Endowment. \\
\raggedright Proceedings of the VLDB Endowment, Vol. \vldbvolume, No. \vldbissue\ %
ISSN 2150-8097. \\
\href{https://doi.org/\vldbdoi}{doi:\vldbdoi} \\
}\addtocounter{footnote}{-1}\endgroup

\ifdefempty{\vldbavailabilityurl}{}{
\vspace{.3cm}
\begingroup\small\noindent\raggedright\textbf{PVLDB Artifact Availability: }\\
The source code, data, and/or other artifacts have been made available at \url{\vldbavailabilityurl}.
\endgroup
}

\section{Introduction}
In recent years, the exponential growth in data volumes~\cite{statistaWorldwideDataCreated} has significantly increased the demands on \gls{dbms} resources and intensified the challenges faced by \glspl{dba}. 
As complex, specialized software designed to manage data efficiently, \gls{dbms} requires substantial computational resources (e.g., CPU, RAM) and expert administration. 
Among these administrative tasks, database tuning is critical, as configuration adjustments directly influence system performance~\cite{tuning}. 
Optimal performance hinges on precise knob tuning, where misconfigurations can lead to significant performance degradation.
Therefore, ensuring that each knob is appropriately tuned is not only essential for maintaining system efficiency but also for achieving reliable and scalable operations.

Modern \gls{dbms}, such as \textsc{MySQL}~\cite{mysql}, expose hundreds of configurable knobs, resulting in a vast tuning space that is both difficult for humans to comprehend and impractical to explore exhaustively. 
Importantly, the influence of these knobs on performance varies---some can have a significant impact, while others contribute minimally---and their importance is often contingent on the specific workload and hardware configuration.
This variability makes knob selection equally critical: identifying and focusing on the most influential knobs can dramatically reduce the search space, improve convergence rates, and lower computational overhead for tuning algorithms.
In the work of~\cite{smac_dbms}, it is demonstrated that the large number of knobs to tune, once they surpass a given threshold, will no longer yield better performance but only to increase the computational overhead.
Consequently, most state-of-the-art \gls{dbms} tuning algorithms limit the tuning scope by selecting only the 10 to 20 most influential knobs~\cite{Hunter,UDOtune,CDBTune}.
%
Emphasizing both knob tuning and informed knob selection is therefore fundamental to achieving efficient and effective \gls{dbms} performance optimization.

Many previous studies have tackled knob selection and \gls{dbms} tuning using a variety of approaches that fall into four broad categories: experience-based, screening-based, ML-based, and \gls{llm}-based methods.
Experience-based systems, like \textsc{CDBTune}~\cite{CDBTune}, use expert-driven knob selection to shrink the search space to apply \gls{ddpg}~\cite{ddpg}, UDO~\cite{UDOtune} extends this with index tuning via \emph{Monte Carlo Tree Search}, but still leaves knob choice to the user.
Screening-based approaches, such as Plackett \& Burman~\cite{sard} and sensitivity analysis~\cite{ituned}, collect knobs-performance samples and apply statistical methods to rank the knob importance.
ML-based approaches exemplified by \textsc{OtterTune}~\cite{Ottertune}, which applies Lasso~\cite{lasso} to prune knobs before \gls{bo}, and \textsc{Hunter}~\cite{Hunter}, which uses Random Forest with \emph{Classification And Regression Trees} (CART)~\cite{cart}--automatically ranking and eliminating low-importance knobs.
Recent \gls{llm}-based methods, such as DB-BERT~\cite{DB_bert} and \textsc{GPTuner}~\cite{GPTuner}, leverage models, like BERT~\cite{bert} and \textsc{ChatGPT}~\cite{ChatGPT}, to parse documentation for knob importance, then hand off tuning to algorithms---like DQN~\cite{double_Q} or SMAC~\cite{SMAC}.

Although these techniques effectively reduce dimensionality, they often incur complex data collection processes and significant computational overhead.
ML-based approaches require extensive pre-collected database performance samples (configuration-to-performance mappings) for model training and knob ranking, making them highly resource-intensive. 
For example, OLTP workload sample collection involves stress-testing the \gls{dbms} under various configurations and computing the average \gls{tps}, whereas OLAP workloads typically require sequential execution of analytical queries, measuring their total run-time under each configuration.
In practice, higher sample counts yield better precision in importance estimates and increase statistical confidence in the rankings.
Although \textsc{Hunter}~\cite{Hunter} reported that collecting only two samples per knob was sufficient for knob ranking, conventional heuristics recommend gathering ten samples per knob to ensure model validity~\cite{peduzzi1996}.
In a DBMS tuning study~\cite{smac_dbms}, more than fifteen samples per knob were used to construct a reliable ranking.
Consequently, accurately ordering a large set of knobs can demand hundreds of measurements, prolonging the sample‐collection and knob selection phase.
Screening methods, like Plackett \& Burman~\cite{sard} and sensitivity analysis~\cite{ituned}, are much more lightweight---typically requiring on the order of twice the number of knobs in experimental runs---yet their importance scores remain fairly coarse, since they do not capture interactions or the full performance-configuration surface.
%

Although experience-based methods leverage human judgment, they are not without their own challenges.
According to~\cite{MYWORK}, even most experienced DB experts struggle with \gls{dbms} tuning, therefore significant time and effort is required to accurately identify and rank the most critical knobs.
%
Lastly, \gls{llm}-based methods are also associated with significant costs and complexities. 
Document collection and organization can be cumbersome, particularly given variations across \gls{dbms} versions and types. 
Furthermore, even after documentation compilation, interactions with language models remain time-consuming.
\textsc{GPTuner}~\cite{GPTuner}, for example, reported more than five hours spent solely on \gls{llm}-based knob selection, accompanied by financial overhead exceeding \$50 in API consumption per tuning.
In summary, current knob selection strategies impose substantial overheads that hinder their widespread adoption.
Together, these requirements in time, compute, and specialized expertise place current methods out of reach for users with limited resources or domain familiarity, highlighting the need for a more efficient and accessible knob‑selection solution.

To bypass complex configurations and extensive data requirements, we introduce a lightweight solution with minimal overhead: \textsf{DOT} (\emph{Dynamic knob selection and Online sampling for automated database Tuning}). 
The core concept of \textsf{DOT} is straightforward: during \gls{dbms} optimization, we continuously collect performance samples and fully exploit them to rank knob importance, prune low-impact knobs, and introduce new knobs dynamically, rather than pre-ranking knobs before tuning.
By dynamically controlling the search‐space dimension through online sampling, \textsf{DOT} focuses only on the most impactful knobs.
Specifically, it applies \gls{rfecv} to prune low‐importance knobs based on samples collected during tuning, while an \gls{lrt} strategy determines whether to explore by adding knobs or to exploit by refining the current set.
To further reduce benchmarking overhead, \textsf{DOT} incorporates an adaptive benchmarking mechanism that accelerates tuning and lowers resource consumption.
For efficient parameter search, \textsf{DOT} employs \gls{bo} to optimize \gls{dbms} configurations on‐the‐fly, eliminating costly warm‐up phases and prior knob‐importance knowledge. Additionally, when prior knob‐importance information is available, \textsf{DOT} can seamlessly integrate it to achieve even faster tuning.


\textbf{Contributions} In this paper, we share the following contributions:
\textbf{(1)} We introduce an experimentation protocol that improves the reproducibility of DBMS tuning algorithms in noisy, stochastic tuning settings;
\textbf{(2)} We conduct an exhaustive empirical study of knob selection in DBMS tuning, revealing the inherent complexity of knob selection and its dependence on workload specificities that motivates \textsc{DOT};
\textbf{(3)} Unlike recent DBMS tuners that rely on extensive pretraining~\cite{Ottertune, CDBTune, Hunter} or heavy LLM‐based documentation analysis~\cite{DB_bert, GPTuner}, we present \textsc{DOT}, a lightweight tuning framework that requires no dedicated warm-up yet can seamlessly incorporate prior knob‐importance information when available; and
\textbf{(4)} Experimental results demonstrate that \textsf{DOT} matches or outperforms state‐of‐the‐art tuners while substantially reducing tuning overhead, with or without prior knob‐importance knowledge.

The paper is organized as follows:
Section~\ref{sec:reprod_protocol} analyzes reproducibility challenges and details our robust evaluation protocol.
Then, Section~\ref{sec:experiments_knob_selection} presents the exhaustive knob‐selection experiments that motivate \textsc{DOT}.
Finally, Section~\ref{DOT_overview} describes the design and operation of \textsc{DOT} and Section~\ref{sec:baselines_and_experiments} evaluates its performance and efficiency against baselines and leading algorithms.

\section{Reproducibility \& Evaluation Challenges in DBMS Auto-tuning}
\label{sec:reprod_protocol}
In this section, we highlight reproducibility challenges in DBMS tuning, demonstrate the high variance in both algorithms and performance measurements, and introduce a robust evaluation protocol for our following experiments.

State-of-the-art \gls{dbms} auto-tuning methods promise significant performance improvements, yet their reproducibility remains challenging due to inherent complexities and environmental variability.
Recent state-of-the-art approaches rely heavily on extensive pre-training and substantial preliminary datasets, which drastically reduce tuning times but introduce reproducibility barriers:
\begin{itemize}[leftmargin=*]
    \item \textbf{Heavy data requirements:} Pre-training typically requires large, high-quality datasets that are rarely shared publicly and costly to collect~\cite{CDBTune,Ottertune}.
    \item \textbf{System Diversity:} Variations in hardware, system configurations, and data collection methods frequently impede accurate reproduction of published results.
    \item \textbf{Complex setup:} Configuring and warm-starting advanced tuning algorithms, such as \gls{drl} like DDPG, demand specialized expertise and hyper-parameter tuning of the algorithm itself~\cite{rl_tuning}.
\end{itemize}

Specific systems exhibit different degrees of these challenges.
For instance, \textsc{CDBTune} requires substantial DBA expertise alongside thousands of high-quality samples; \textsc{Hunter} mandates dedicated infrastructure to generate extensive initial datasets; \textsc{OtterTune} relies on large-scale metric aggregation pipelines; \emph{UDO} demands manual selection of knobs by users; and recent \gls{llm}-based methods (e.g., \emph{DB-BERT}, \textsc{GPTuner}) hinge on extensive documentation parsing, dependent on the capabilities of the underlying LLM.
Consequently, setup complexity and data collection overhead inflate overall tuning duration, impair reproducibility, and limit
applicability.

In contrast, \textsf{DOT} adopts a lightweight strategy, dispensing entirely with pre-training, thus minimizing preliminary data collection and expert intervention. This significantly enhances reproducibility and broadens potential industrial adoption.

Beyond reproducibility barriers inherent to current tuning strategies, another critical challenge is the variability inherent to noisy DBMS tuning environments.
Performance measurements of DBMS benchmarks display substantial noise due to factors such as transient network congestion~\cite{network1,network2}, OS/VM background activities~\cite{os1}, intrinsic benchmark randomness~\cite{sysbench,tpcc}, and hardware side-effects like CPU throttling~\cite{CPU1,CPU2}.
We evaluated performance variability across three representative workloads (two OLTP workloads: TPC-C and \textsc{Sysbench}, and one OLAP workload: TPC-H).
Table~\ref{tab:variation_benchmark} quantifies run-to-run variation for 10 runs on the same \textsc{MySQL} database under default configuration.
\setlength{\textfloatsep}{6pt}
\begin{table}[htbp]
\caption{Variability of performance measurements}
\label{tab:variation_benchmark}
\centering
\footnotesize
\begin{tabular}{lc|rrr}
\toprule
\textbf{Benchmark} & \textbf{Type} & \textbf{Mean}  & \textbf{SD} & \textbf{CV (\%)} \\
\midrule
TPC-H    & OLAP &  80.13 sec. &  0.42 & 0.52 \\
TPC-C    & OLTP &  79.33 TPS  &  4.95 & 6.24 \\
\textsc{Sysbench} & OLTP & 608.63 TPS  & 26.53 & 4.36 \\
\bottomrule
\end{tabular}
\end{table}

For OLAP workloads, we use the total execution time as a measurement of their performance, while for OLTP workloads, we use the TPS as a performance metric. 
It is observed that even under an \emph{identical} configuration, the coefficient of variation (CV\(=\frac{\text{SD}\times100}{\text{Mean}}\)) reaches $6.2\%$ for TPC-C and $4.4\%$ for \textsc{Sysbench}.
TPC-H presents a relatively lower variance of $0.52\%$.
%
Additionally, ML-based tuning algorithms, such as \emph{Bayesian Optimization} (BO), exhibit inherent stochasticity~\cite{AI_reproducibility}, leading to considerable variability in tuning speed and convergence behavior.
To systematically assess convergence, we define a \emph{near-optimality} criterion based on the workload type.

\textbf{For OLTP workloads}, we declare convergence once throughput stabilizes within noise. 
Formally, let $t^*_{\text{OLTP}}$ be the first iteration (after 100 steps) where the best throughput over the past 100 steps improves by less than $5\%$. 
The optimal throughput is 
$
P^{*} = \max_{t < t^{*}_{\text{OLTP}}}\text{max\_tps}(t),
$
and convergence occurs at
$
Y_{\text{OLTP}} = \min\{t \mid \text{max\_tps}(t) \ge 0.95\,P^{*}\},
$
i.e., when throughput reaches 95\% of $P^{*}$ (matching the $\sim$5\% noise level).

\textbf{For OLAP workloads}, we apply the same idea to query-batch execution time. 
Here $t^*_{\text{OLAP}}$ is the first iteration (after 100 steps) where the best time improves by less than $1\%$, reflecting the lower noise of OLAP benchmarks. 
The optimal execution time is 
$
E^{*} = \min_{t < t^{*}_{\text{OLAP}}}\text{total\_exec\_time}(t),
$
and convergence is reached at
$
Y_{\text{OLAP}} = \min\{t \mid \text{total\_exec\_time}(t) \le 1.01\,E^{*}\}.
$

Table~\ref{tab:variation_algo} highlights the inherent stochasticity of ML-based optimizers, illustrating significant variability in iterations and time to reach convergence.
The number of iterations required to reach near-optimality varies sharply: TPC-C shows the widest absolute spread (standard deviation of about 122 iterations), whereas TPC-H is the noisiest in relative terms (coefficient of variation of about 73\%), confirming that convergence speed can differ markedly from run to run.
The time needed to achieve near optimality exhibits similar volatility. 
\gls{bo} needs 3.4\,h for \textsc{Sysbench}, 6.9\,h for TPC-C, and 1.2\,h for TPC-H, on average, but the corresponding standard deviations (1.2\,h, 4.4\,h, and 0.9\,h) translate into coefficients of variation of 35\%, 64\%, and 73\%, respectively. 

\begin{table}[h!]
    \centering
    \footnotesize
    \caption{Convergence variability of the BO tuner: iterations \& time to reach near-optimal performance across benchmarks}
    \label{tab:variation_algo}
    \begin{tabular}{l|ccc|ccc}
    \toprule
    \multirow{2}{*}{\textbf{Benchmark}} &
      \multicolumn{3}{c|}{\textbf{Iterations}} &
      \multicolumn{3}{c}{\textbf{Time (hours)}} \\
    \cline{2-7}
             &  Mean &    SD & CV\,(\%) & Mean & SD & CV\,(\%) \\
    \midrule
    TPC-H    &  57.0 &  41.3 & 72.5 & 1.21 & 0.9 & 72.9 \\
    TPC-C    & 215.6 & 122.2 & 56.7 &  6.9 & 4.4 & 64.3 \\
    \textsc{Sysbench} & 118.4 &  40.3 & 34.1 &  3.4 & 1.2 & 34.6 \\
    \bottomrule
    \end{tabular}
\end{table}

Given the substantial noise introduced by both the execution environment and the tuning algorithms, we adopt a more rigorous evaluation protocol to mitigate bias:
\begin{enumerate}[leftmargin=*]
  \item \textbf{Replication:} We execute each tuning process \emph{five} times with distinct random seeds, performing a full DBMS reboot between runs to minimize warm‐up and caching artifacts.
  \item \textbf{Aggregate reporting:} results are summarized by the sample mean and standard deviation across all trials, with 95\% confidence intervals shown in our figures to convey uncertainty.
  \item \textbf{Statistical testing:} We assess the significance of performance differences using 
Welch’s two-tailed \(t\)-test~\cite{welch} for pairwise comparisons and the non-parametric 
Friedman test~\cite{friedman} for multiple comparisons. Both tests rely on their corresponding 
\(p\)-values to quantify the likelihood of observing the reported differences under the null 
hypothesis. We adopt a significance threshold of \(p < 0.05\).

\end{enumerate}

While this protocol reduces bias, it cannot eliminate it entirely.
In principle, a larger number of repetitions would yield more definitive comparisons, but each full tuning run requires two machines for over 10 hours, making extensive replication prohibitively expensive; hence, we settle on five repetitions in this study.

\section{Experimental Study of Knob Selection \& Challenges}
\label{sec:experiments_knob_selection}

In this section, we analyse different knob selection strategies and present experimental evidence that underscores the inherent challenges of knob selection in DBMS tuning.  
These difficulties motivate the need for a dynamic, efficient, and cost-aware approach—precisely what our proposed method, \textsf{DOT}, is designed to deliver.


Two OLTP workloads (Sysbench with 10 tables of 2 million rows and 50 concurrent threads, implementation from~\cite{sysbench}; TPC-C at scale 100 with 32 threads, implementation from~\cite{tpcc_imp}) and one OLAP workload (TPC-H at scale 1, implementation from~\cite{tpch_imp}) are used for the following evaluations.
Experiments were conducted on \textsc{MySQL}\,8 running on a VM (4\,vCPU, 16\,GB RAM) in Orange’s FlexEngine~\cite{orangeFlexibleEngine} cloud platform.

\subsection{Analysis of Knob Selection Strategies}
\label{sec:experiments_knob_selection_strategies}

\begin{table}[h!]
  \centering
  \caption{Performance tuned with TOP-5 knobs selected by different methods}
  \label{tab:top5_perf_methods}
  \resizebox{\columnwidth}{!}{%
    \begin{tabular}{lrrrrrr}
      \toprule
      Workload      & Sensitivity   & P \& B  &  Lasso   & CART    & Experience & LLM \\
      \midrule
      Sysbench (TPS)      &  1411.1 &  939.1  & 1469.04  & \textcolor{red}{1588.7} & 1294.36    & 1411.1     \\
      TPC-H (Exec. s)     &   46.22 &   74.35 &   46.02  &    45.76                & \textcolor{red}{45.72} &  46.35     \\
      \bottomrule
    \end{tabular}%
  }
\end{table}

First, we evaluate the performance of different knob-selection strategies by comparing six methods on a fixed set of 22 expert-identified knobs using two representative benchmarks: Sysbench (OLTP) and TPC-H (OLAP).

Each method—P\&B~\cite{sard}, Lasso~\cite{Ottertune}, CART~\cite{Hunter}, Sensitivity Analysis~\cite{ituned}, an experience-based approach, and LLM knowledge based on \textsc{GPTuner}’s prompt\cite{GPTuner} to ask \textsc{ChatGPT}\,4o to produce a ranking of the 22 knobs.
We collected 48 configuration–performance samples for P\&B, 220 samples each for CART and Lasso, and 44 samples for Sensitivity Analysis, each following their respective ranking protocols.
In contrast, \textsc{ChatGPT}\,4o required no sampling, significantly reducing the overhead of knob ranking and the experience-based approach requires input of a senior \gls{dba}.

Subsequently, we conducted an exhaustive grid search over the top five knobs identified by each method, evaluating integer knobs at five discrete values and boolean knobs at two states, generating up to $5^5 = 3,125$ configurations per knob set (approximately 78 hours of benchmarking each). 
Due to the computational burden, we sampled the performance–configuration space only once (despite benchmarking noise) and temporarily suspended the five-repetition protocol, accepting the resulting increase in variance as an unavoidable trade-off to make this small-scale comparison.

Grid search ensures unbiased evaluation by eliminating tuning algorithm variability, allowing a pure assessment of knob selection methods.
Although real-world configurations often include many more knobs, exhaustively searching them would incur exponential computational costs; therefore, we limit our study to five.
Table~\ref{tab:top5_perf_methods} compares the performance under different knob sets:
\begin{enumerate}[leftmargin=*]
    \item \textbf{CART} achieves the highest {\sc Sysbench} throughput (1,588.7~TPS) and among the lowest TPC-H execution times (45.76~s), indicating superior overall knob-selection effectiveness. However, CART requires the most extensive data collection (220 samples).
    \item \textbf{Lasso} and \textbf{LLM knowledge} match CART at specific points in {\sc Sysbench} tuning but underperform for OLAP.
    Lasso has a high sampling cost as CART, whereas \textsc{ChatGPT} achieves comparable rankings with low overhead (a single prompt).
    \item \textbf{Experience-based} and \textbf{Sensitivity Analysis} methods yield moderate improvements over default settings but do not reach CART’s best performance.
    \item \textbf{P\& B} demonstrates modest gains with the least sampling requirements, reflecting a lower resource–performance trade-off.
\end{enumerate}

These preliminary findings—derived from single exhaustive grid searches per configuration—demonstrate CART’s strong knob selection performance alongside its substantial sampling overhead, clearly illustrating the trade-off between sampling cost and performance gains and underscoring that sub-optimal knob choices directly impair tuning results.
As initial guidance, they highlight the challenge of balancing sampling effort against tuning efficacy.

\begin{figure}
    \centering
     \includegraphics[
    width=0.8\linewidth,
    trim=0   1cm   0   0,
    clip
  ]{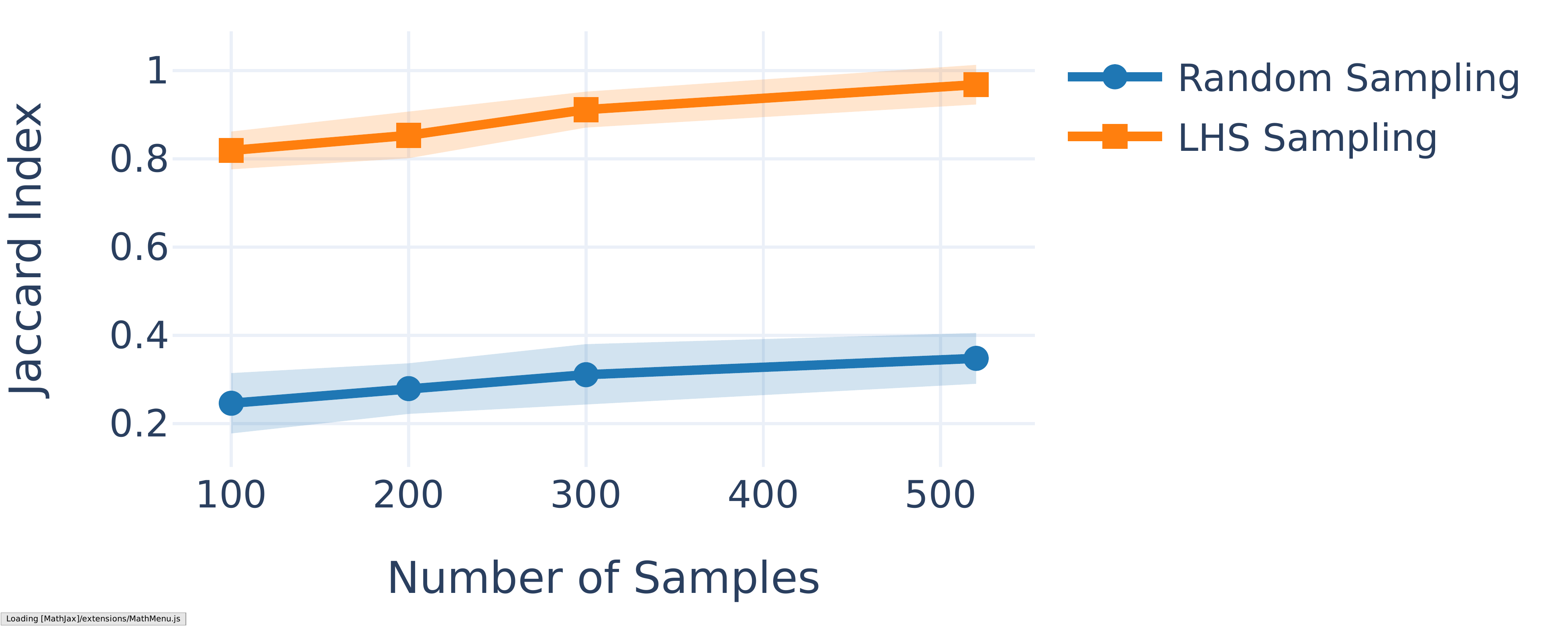}
    \caption{Variability of CART knob selection across varying sample sizes \& sampling strategies}
    \label{fig:Jaccard}
\end{figure}

\subsection{Cost of Robust Knob Rankings}
To further investigate the cost of robust knob ranking, we adopt a more practical setting to knob selection and DBMS tuning by moving beyond the narrow 22-knob subset. 
Recognizing that many \textsc{MySQL} variables govern security, file paths, or other non-performance concerns—and constrained by our compute budget—we filtered the full set of tunable parameters down to the 52 most performance-relevant knobs (primarily InnoDB settings such as "innodb\_purge \_threads", "innodb\_read\_io\_threads", "innodb\_buffer\_pool\_size").  
More details of the experimental setup and the complete knob lists and configuration scripts are available in the \hyperlink{avail}{Availability statement}.

This targeted reduction both removes irrelevant options and keeps the global optimization search space realistic. 

We delve deeper into the computational expense and variability of the CART method from \textsc{Hunter}, which constructs a Random Forest and aggregates knob importance via majority voting among CART trees~\cite{Hunter}.
Given that ML algorithms such as CART are inherently stochastic, their knob rankings may vary with different random seeds, making reproducibility an important concern.

To quantify this variability, we use CART to rank the pre-selected 52 knobs, choosing the Top 20 as in \textsc{Hunter}~\cite{Hunter}. 
We train CART models with varying sample sizes and random seeds, then compute the Jaccard index between each pair of Top\,20 sets to measure how consistently the same knobs are selected. 
The Jaccard index is defined as $J(A, B) = \dfrac{\lvert A \cap B \rvert}{\lvert A \cup B \rvert}$, where $J(A, B)\in [0,1]$, with higher values indicating greater similarity between two sets.

Different sampling strategies may also influence the results. 
We compare \emph{Latin Hypercube Sampling} (LHS)---used by \textsc{OtterTune}~\cite{Ottertune} and \textsc{Hunter}~\cite{Hunter}---against pure random sampling (as in \textsc{CDBTune}~\cite{CDBTune}).
Figure~\ref{fig:Jaccard} plots Jaccard indices for the \textsc{Sysbench} workload on \textsc{MySQL} when using 100, 200, 300, and 520 samples (520 aligns with ten samples per feature~\cite{peduzzi1996}).
Under random sampling, Jaccard indices remain low: different seeds yield very different Top\,20 sets, indicating high stochasticity.
In contrast, LHS produces higher Jaccard indices and more consistent knob selections.
In both cases, increasing the number of samples improves stability.
However, collecting samples is costly: each sample requires one client machine running \textsc{Sysbench} against a DBMS server for about 100 seconds, so 520 samples take roughly 14.4 hours on two machines.

Additionally, we employ \gls{bo} (implemented using Scikit Optimize~\cite{scikitoptimize}) to tune the DBMS with the Top\,20 knobs identified through CART rankings derived from different numbers of samples generated by LHS.
Although \gls{bo} inherently introduces variability, grid search becomes infeasible due to the relatively high dimensionality (20 knobs), making \gls{bo} a more practical choice.
Following the protocol detailed in Section~\ref{sec:reprod_protocol}, we record the final throughput and total tuning time to evaluate how effectively each ranking identifies the most impactful knobs.

Intuitively, providing \gls{CART} with more samples enhances knob‐ranking quality by supplying richer information.
Table~\ref{tab:samples} presents the tuned DBMS performance, Friedman test rankings, and sampling times using knob sets selected via CART at varying sample sizes via \gls{bo}. 
Generally, as the sample size increases, tuned DBMS performance does improve.
However, performance gains beyond 200 samples become marginal and statistically insignificant according to the t-test.
Specifically, increasing from 100 to 200 samples notably enhances performance (from 1,602.3~TPS to 1,690.8~TPS), but further increments (e.g., from 200 to 300 or 520 samples) offer limited benefits.

Regarding tuning time, results indicate no clear correlation with sample size, as tuning duration fluctuates without a distinct pattern.
Importantly, sampling time significantly surpasses tuning time in all cases, dominating the total effort.
Sampling 100 configurations already requires approximately 167 minutes, rising sharply to about 333 minutes (over 5 hours) for 200 samples, and further ballooning to approximately 867 minutes (over 14 hours) for 520 samples.

These findings underscore a crucial trade-off: increasing sample sizes provides diminishing returns in performance improvement but incurs growing sampling costs.
Our experiments suggest that selecting around 200 samples achieves an optimal balance between improved performance and manageable sampling effort, although this optimal point could vary with different workloads and scenarios.
Thus, careful consideration of this trade-off is essential, as proper and robust knob selection methods are inherently costly.

\begin{table}[htbp]
  \caption{Sysbench tuning with CART on knob sets from varying sample sizes (statistically significant at $p <$ 5E-2)}
  \label{tab:samples}
  \centering
  \resizebox{\columnwidth}{!}{%
    \begin{tabular}{@{}c|ccc@{}}
      \toprule
     \textbf{Samples} & \textbf{Rank (Max TPS)} & \textbf{Rank (tuning time [min])} & \textbf{Sampling} \\
        & Test stat:~$7.32$ 
        & Test stat:~$1.32$ 
        & \textbf{time (min)} \\
        & p‐value:~$6.2E-2$ 
        & p‐value:~$7.2E-1$ 
        &  \\
      \midrule
      520  
        & \textcolor{red}{$1.8$} (1{,}708.0~$\pm$~29.1) 
        & $2.8$ (216.2~$\pm$~32.6) 
        & 866,7  \\
      300     
        & $2.0$ (1{,}684.3~$\pm$~48.9) 
        & \textcolor{red}{$2.0$} (160.8~$\pm$~79.5) 
        & 500,0  \\
      200     
        & $2.4$ (1{,}690.8~$\pm$~65.2) 
        & $2.8$ (214.9~$\pm$~116.9) 
        & 333.3  \\ 
      100       
        & $3.8$ (1{,}602.3~$\pm$~59.5)                           
        & $2.4$ (181.3~$\pm$~36.1)   
        & 166.7  \\ 
      \bottomrule
    \end{tabular}}
\end{table}

\subsection{Workload-Sensitive Variations of Top Knobs}
Additionally, knob importance can vary dramatically between workloads. 
To demonstrate this, we drew 520 samples each via LHS and used CART to pick the Top\,20 knobs (out of 52) for \textsc{Sysbench}, TPC-C, TPC-H workload. 
We then computed the Jaccard similarity between each pair of Top\,20 sets and a general Top\,20 knobs selected by expertise without workload specificity (so-called EXP) and presented in Table~\ref{tab:jaccard-matrix}.
Overall, these results reveal relatively low overlap among workloads and expert rankings.
Contrary to our expectation that the two OLTP workloads (TPC-C and \textsc{Sysbench}) would overlap most, TPC-C actually shares more knobs with OLAP workload (TPC-H). 

\begin{table}[h!]
\centering
\caption{Jaccard index between knob sets.}\label{tab:jaccard-matrix}
\resizebox{\columnwidth}{!}{%
\begin{tabular}{l|cccc}
\toprule
            & \textsc{Sysbench}          & TPC-C             & TPC-H             & EXP                \\ 
\midrule
\textsc{Sysbench}    & —                 & $0.317 \pm 0.035$ & $0.325 \pm 0.033$ & $0.258 \pm 0.016$  \\
TPC-C       & $0.317 \pm 0.035$ & —                 & $0.472 \pm 0.041$ & $0.243 \pm 0.029$  \\
TPC-H       & $0.325 \pm 0.033$ & $0.472 \pm 0.041$ & —                 & $0.409 \pm 0.024$  \\
EXP         & $0.258 \pm 0.016$ & $0.243 \pm 0.029$ & $0.409 \pm 0.024$ & —                  \\
\bottomrule
\end{tabular}
}
\end{table}


These findings show that a precise knob selection must be repeated for each workload—and if that step is too costly, it can bottleneck any \gls{dbms} auto-tuning approach. Yet we also uncovered a surprising degree of transferability. Table~\ref{tab:merged_sysbench} compares \textsc{Sysbench} throughput when tuned with 6 different knob sets.
Specifically, we apply the top 20 knobs identified for each workload to optimize the \textsc{Sysbench} benchmark using \gls{bo}:
\begin{itemize}[leftmargin=*]
  \item \textbf{Sysbench/TPC-C/TPC-H}: Top\,20 knobs ranked by CART for the corresponding workload (520 LHS samples).
  \item \textbf{Exp}: Top\,20 knobs ranked by experts.
  \item \textbf{Shared}: the 7 knobs appearing in all three Top\,20 lists.
  \item \textbf{Random}: 20 knobs chosen at random from the full set.
\end{itemize}

\begin{table}[t!]
  \caption{\textsc{Sysbench} tuning with knob sets for different workloads ranked with CART (statistically significant at $p <$ 5E-2)}
  \label{tab:merged_sysbench}
  \centering
  \footnotesize
    \begin{tabular}{@{}l|cc@{}}
      \toprule
      & \textbf{Rank (Max TPS)}      & \textbf{Rank (tuning time [min])}   \\
      & Test stat: 12.7, p‐value: 2.7E‐2  & Test stat: 17.8, p‐value: 3E‐3   \\
      \midrule
      \textsc{Sysbench} 
        &   $2.4$ (1{,}708.0~$\pm$~29.1) 
        & \textcolor{red}{$2.4$} (216.2~$\pm$~32.6) \\
      TPC‐C    
        & $4.0$ (1{,}653.2~$\pm$~27.2) 
        & $3.0$ (218.4~$\pm$~65.3) \\
      TPC‐H    
        & $4.0$ (1{,}638.8~$\pm$~35.2) 
        & $2.8$ (188.3~$\pm$~57.9) \\ 
      EXP      
        & \textcolor{red}{$1.4$} (1{,}725.2~$\pm$~31.1)                           
        & $2.6$ (217.8~$\pm$~42.7)                           \\
      Shared   
        & $3.2$ (1{,}677.1~$\pm$~35.2) 
        & $5.8$ (447.3~$\pm$~26.2) \\
      Random   
        & $6.0$ (1{,}063.1~$\pm$~82.9) 
        & $4.4$ (308.8~$\pm$~169.8) \\
      \bottomrule
    \end{tabular}%
\end{table}

To our surprise, the expert-ranked knob set (EXP) achieves the highest throughput ($1{,}725\pm31$ TPS), while the \textsc{Sysbench}-specific knobs converge slightly faster ($216.2\pm32.6$ min vs.\ $217.8\pm42.7$ min). 
However, their performance and tuning time are statistically equivalent according to the t-test---indicating that both represent strong, practical choices.
Tuning the \textsc{Sysbench} knobs delivers excellent performance ($1{,}708\pm29$ TPS), reinforcing the effectiveness of workload-specific ranking, while the EXP result highlights the potential of domain knowledge to generalize well.

Other workload-derived knob sets remain competitive: using the TPC‑C and TPC‑H rankings yields $1{,}653\pm27$ TPS and $1{,}639\pm35$ TPS, close to the expert-ranked result.
Even the shared important knob set (7 knobs common to all workloads) remains close to the expert-level throughput ($1{,}677\pm35$ TPS).
In contrast, random knob selection performs poorly, losing nearly 40\% of the attainable throughput ($1{,}063\pm83$ TPS).
For tuning time, the TPC-H knob set even outperforms others, reaching convergence in just $188.3\pm57.9$ minutes.
However, restricting the search to shared knobs doubles the convergence time ($447.3\pm26.2$ min), indicating that omitting workload-specific knobs can hinder optimizer guidance.
Random selection also increases tuning duration ($308.8\pm169.8$ min), despite the performance degradation.

While the knob sets derived from different workloads exhibit relatively low overlap, some of them achieve similar performance when transferred. 
This suggests that, despite divergent rankings, multiple knob subsets may still expose sufficiently informative dimensions of the configuration space for effective tuning.
To summarize: 
\textbf{1) Cross‑workload reuse is viable}: top knobs from a workload can be reused in another with modest loss in throughput.
\textbf{2) Common knobs matter, but are not sufficient}: restricting to universally important knobs retains much of the performance but incurs convergence delays.
\textbf{3) Expert knowledge or workload-aware rankings remain best}: they yield the highest throughput and efficient tuning.


In practice, prior knowledge or expert insight can provide a useful warm start, skipping a full ranking phase.
However, blindly reusing outdated or irrelevant knob sets risks missing critical parameters, leading to poor performance or longer tuning times.



\subsection{Optimal Number of Tuned Knobs} 
Additionally, choosing how many knobs to tune is nontrivial: too few knobs may yield poor performance, while too many inflate overhead and extend search time.
\begin{table*}[ht]
  \caption{Performance, tuning time, and Friedman‐test ranks for different numbers of knobs (statistically significant at p < 5e-2)}
  \label{tab:merged_final}
  \centering
  \scriptsize
  \resizebox{\textwidth}{!}{%
    \begin{tabular}{@{}l|cc|cc|cc@{}}
      \toprule
      & \multicolumn{2}{c|}{\bfseries \textsc{Sysbench}}
      & \multicolumn{2}{c|}{\bfseries TPC-C}
      & \multicolumn{2}{c}{\bfseries TPC-H} \\
      \cmidrule(lr){2-3}\cmidrule(lr){4-5}\cmidrule(l){6-7}
      & \begin{tabular}{@{}c@{}}Rank (Max TPS)\\Test stat: 13.3\\p-value: 1E-2\end{tabular}
      & \begin{tabular}{@{}c@{}}Rank (tuning time [mins])\\Test stat: 9.1\\p-value: 5.8E-2\end{tabular}
      & \begin{tabular}{@{}c@{}}Rank (Max TPS)\\Test stat: 12.0\\p-value: 1.7E-2\end{tabular}
      & \begin{tabular}{@{}c@{}}Rank (tuning time [mins])\\Test stat: 5.1\\p-value: 2.8E-1\end{tabular}
      & \begin{tabular}{@{}c@{}}Rank (Exec (s))\\Test stat: 6.1\\p-value: 1.9E-1\end{tabular}
      & \begin{tabular}{@{}c@{}}Rank (tuning time [mins])\\Test stat: 11.8\\p-value: 1.9E-2\end{tabular} \\
      \midrule
      Top\,5 knobs  
        & 4.6 (1,486.8 ± 159.0)  
        & 3.0 (311.5 ± 194.2)  
        & 5.0 (120.5 ± 22.1)  
        & 2.4 (151.7 ± 161.1)  
        & 3.6 (47.7 ± 0.3)  
        & \textcolor{red}{1.6} (46.1 ± 54.5) \\
      Top\,10 knobs 
        & 4.2 (1,674.9 ± 34.4)  
        & 2.6 (272.0 ± 122.8)  
        & 3.2 (338.5 ± 13.8)  
        & \textcolor{red}{2.0} (153.8 ± 186.7)  
        & 3.8 (47.7 ± 0.5)  
        & 2.0 (32.7 ± 14.6) \\
      Top\,20 knobs 
        & \textcolor{red}{2.0} (1,725.2 ± 31.1)  
        & \textcolor{red}{2.0} (217.8 ± 42.7)  
        & \textcolor{red}{1.8} (354.2 ± 14.8)  
        & 3.4 (411.9 ± 264.9)  
        & 3.2 (47.8 ± 1.4)  
        & 3.2 (73.0 ± 53.2) \\
      Top\,30 knobs 
        & \textcolor{red}{2.0} (1,728.4 ± 46.3)  
        & 2.6 (267.9 ± 136.3)  
        & 2.4 (347.6 ± 25.8)  
        & 3.2 (261.0 ± 287.1)  
        & 2.8 (47.8 ± 2.0)  
        & 3.6 (189.6 ± 286.5) \\
      All\,52 knobs 
        & 2.2 (1,720.0 ± 33.0)  
        & 4.8 (732.6 ± 266.7)  
        & 2.6 (346.0 ± 7.5)   
        & 4.0 (560.4 ± 327.7)  
        & \textcolor{red}{1.6} (46.5 ± 0.6)   
        & 4.6 (334.1 ± 175.5) \\
      \bottomrule
    \end{tabular}%
  }
\end{table*}
Table~\ref{tab:merged_final} reports DBMS performance when tuning different numbers of top knobs (expertise-based) using \gls{bo}.
On \textsc{Sysbench}, tuning only the Top 5 knobs yields lower performance than 10–30 knobs (or all 52) and does not minimize convergence time.
The Top 20 set offers the best trade-off: near-optimal performance (comparable to 30 knobs) with the lowest tuning time.
Although tuning times are statistically similar ($p>0.05$), using all 52 knobs still incurs substantially longer convergence.

For \textsc{TPC-C}, tuning only five knobs performs significantly worse than larger sets.
The Top 20 knobs give the best average performance, while the Top 10 offer the best speed–quality trade-off.
As with \textsc{Sysbench}, tuning all knobs greatly prolongs convergence.

For \textsc{TPC-H}, performance is largely insensitive to knob count ($p>0.05$).
Tuning all knobs yields the best absolute result, but with heavy overhead, making the Top 5 knobs preferable for comparable performance at minimal cost.

These results underscore the challenge of tuning the right number of knobs to balance computational overhead against final performance, and highlight the different behavior for different workloads. 
Moreover, the optimal knob count (whether to minimize tuning time, maximize performance, or achieve both) varies across workloads. 
As a result, identifying a reasonable subset of knobs to balance performance gains against tuning cost often requires repeated experiments, making it difficult to apply in practice.

\subsection{Summary of Knob Selection Challenges}
To summarize, to achieve good \gls{dbms} performance tuning, knob selection is critical.
However, in practice, it raises several challenges:
1)\,\textbf{Identifying the most influential knobs}—poor choices lead to sub-optimal performance and longer tuning cycles, and the cost of important knob identification is high;
2)\,\textbf{Workload dependency}—the set of top-ranked knobs varies by workload, indicating that knob importance depends not only on the \gls{dbms} itself but also on the specific workload characteristics;
3)\,\textbf{Transferability versus specificity}—although top-ranked knobs often transfer reasonably well across workloads, using workload-specific rankings yields better results;
4)\,\textbf{Sampling overhead}—robust knob selection requires extensive sampling to accurately rank parameters, which can be both time-consuming and resource-intensive; and
5)\,\textbf{Choosing how many knobs to tune}—balancing computational overhead, convergence time, and final performance is nontrivial—tuning too few knobs can hurt performance, while tuning too many increases search time without proportional gains. These observations highlight the need and the motivation for a more flexible and dynamic knob selection strategy.

\section{System Overview}
\label{DOT_overview}

\subsection{\textsf{DOT}}

\begin{figure}[t]
    \centering
    \includegraphics[width=0.9\linewidth]{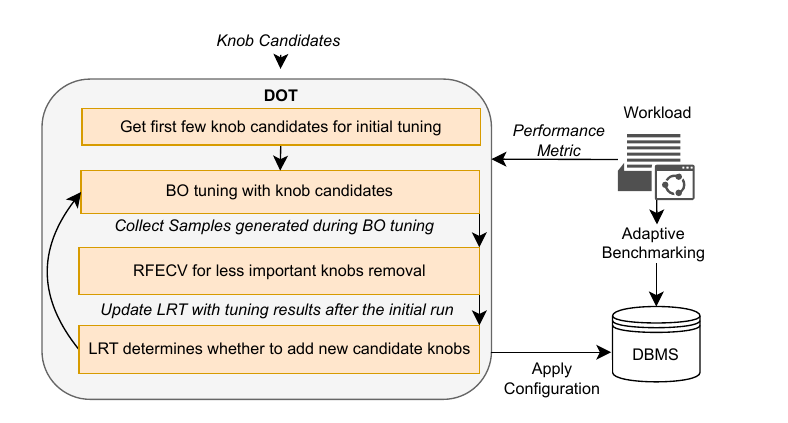}
    \caption{\textsf{DOT} overview}
    \label{fig:DOT_overview}
\end{figure}

To address the above-mentioned practical challenges of knob selection in \gls{dbms} tuning, we present \textsf{DOT}---a dynamic and adaptive framework that continuously adjusts the active knob set during the tuning process for the given workload, focusing on the most impactful knobs as new evidence is collected (Figure~\ref{fig:DOT_overview}).
Unlike traditional methods that perform an up-front knob selection phase, \textsf{DOT} continuously updates the set of knobs being tuned \emph{on the fly}. 
It uses RFECV to assess knob importance from performance samples gathered during the tuning process itself, eliminating the need for costly analysis. 
\gls{bo} is applied to explore high-potential configurations efficiently. 
When available, \textsc{DOT} can use prior knowledge for more efficient tuning.

Algorithm~\ref{alg:DOT-full} describes the procedure. 
Here, \(k_0\) is the initial number of knobs to be tuned; it is set to 20 in our main experiments, following common practice in prior work (\cite{Hunter}) and to ensure fair comparison with baselines. 
Although \(k_0\) is hard-coded during initialization, \textsc{DOT} is relatively insensitive to its specific value (see Section~\ref{sec:ablation}). 
If a ranked knob list is available, \textsf{DOT} tunes knobs in order of importance, beginning with the top $k_0$ knobs. In the absence of such prior knowledge, \textsf{DOT} instead initializes by randomly selecting $k_0$ knobs.
By “sorted list” we mean knobs ordered by an estimated importance (e.g., from expert knowledge, documentation, or ranking heuristics), which allows DOT and other tuning algorithms to prioritize impactful knobs early and avoid wasted trials. 
We distinguish between two practical contexts: (i) when prior knowledge is available (e.g., a knowledgeable user or previously studied workload), DOT and baselines start from this ranked list; (ii) when no such knowledge is available (e.g., less experienced users or unseen workloads), DOT proceeds without ordering and relies entirely on its internal mechanism to discover impactful knobs. 


\gls{bo} with surrogate model Gaussian Process tunes over an epoch with a length of 5 times the number of newly introduced knobs.
During each iteration, we generate one configuration–performance sample $(x,y)$, yielding five samples per knob to ensure stability and reliable knob selection for RFECV (Section~\ref{sec:rfecv}).
After the epoch completes, RFECV is applied to the collected samples to identify the most impactful knobs.
Next, a likelihood‐ratio test (LRT; Section~\ref{sec:lrt-two-action}) decides whether to continue fine‐tuning the reduced knob set or to expand the search space by appending additional knobs.
The LRT’s reward signal is the per‐call performance gain: if exploiting the current set is statistically justified, \textsf{DOT} proceeds with BO on those knobs; otherwise, it explores by adding more knobs.

\begin{algorithm}[H]
\caption{\textsf{DOT}: Dynamic Knob Tuning}
\label{alg:DOT-full}
\begin{algorithmic}[1]
\Require Sorted knob list $\mathcal{K}$; tuning budget $T_{\max}$; initial size $k_0{=}20$; step size $\Delta k{=}5$
\Procedure{DOT}{$\mathcal{K}, T_{\max}, \Delta k$}
  \State $\mathcal{K}_{\text{curr}} \gets \text{top-}k_0(\mathcal{K})$
  \State $\mathcal{D} \gets \varnothing$ \Comment{full-space observations $(x,y)$}
  \State $b \gets -\infty$ \Comment{best performance so far}
  \State $E \gets 5 \cdot k_0$ \Comment{initial epoch length}
  \State $\text{elapsed} \gets 0$
  \While{$\text{elapsed} < T_{\max}$}
    \State \textbf{Step 1: BO Exploration} \Comment{Explore current subspace with BO, record best value and update full observations}
    \State $(\mathcal{D},\, y^*) \gets \Call{BO}{\mathcal{K}_{\text{curr}},\, E,\, \mathcal{D} \to \mathcal{D}_{\text{proj}}}$
    \State \textbf{Step 2: Knob Selection} \Comment{Select most relevant knobs via RFECV in current subspace}
    \State $\mathcal{K}^* \gets \Call{RFECV}{\mathcal{K}_{\text{curr}},\, \mathcal{D} \to \mathcal{D}_{\text{proj}}}$
    \State \textbf{Step 3: Knob Set Expansion} \Comment{Decide shrink/expand using likelihood ratio test}
    \State $a \gets \Call{LRT\_Step}{b,\, y^*,\, E}$; \quad $b \gets \max(b, y^*)$
    \State $\mathcal{K}_{\text{curr}},\, E \gets 
      \begin{cases}
        (\mathcal{K}^*,\, 10), & a=0 \\
        (\mathcal{K}^* \cup \text{next }\Delta k \text{ knobs},\, 5 \cdot \Delta k), & a=1
      \end{cases}$
    \hspace*{3.5em}\Comment{$a=0$: Shrink/Stay, \;\; $a=1$: Expand}
    \State $\text{elapsed} \gets \text{elapsed} + E$
  \EndWhile
  \State \Return best configuration found in $\mathcal{D}$
\EndProcedure
\end{algorithmic}
\end{algorithm}

Additionally, to reduce heavy benchmarking time, we introduce an adaptive benchmark budget to reduce tuning time. 
This process repeats until the overall evaluation budget \(T_{\max}\) is exhausted.
Throughout, \textsf{DOT} reuses prior samples \((x, y) \in \mathcal{D}\) that are projected to the current knob set $\mathcal{D}_{\mathrm{proj}}$, as new knobs are initially set to default values—ensuring that earlier configurations remain valid for initializing BO when the knob set \(\mathcal{K}_{\mathrm{curr}}\) is updated.

This design lets \textsf{DOT} dynamically adjust tuning complexity, balance exploration versus exploitation, and efficiently converge on optimal configurations.
\textsf{DOT}—comprising tuning, assessing knob importance, and refining the search space—follows these principles:

\begin{enumerate}[leftmargin=*]
    \item \textbf{Exploration when impact is limited:} If recent tuning yields minimal improvement, the LRT policy triggers expansion by appending \(\Delta k\) new knobs from the ranked list \(\mathcal{K}\), increasing the tuning space.
    \item \textbf{Exploitation when knobs are effective:} If current knobs drive sufficient gains, RFECV prunes them to retain the influential knobs-- reducing dimensions and improving convergence.
    \item \textbf{Dynamic re-evaluation:} Knob importance is continuously revisited after each epoch, allowing \textsf{DOT} to drop previously selected knobs when more impactful ones emerge.
    \item \textbf{No warm-up required:} \textsf{DOT} starts tuning immediately without a separate knob-selection phase. If prior ranking is available, it can be used to initialize \(\mathcal{K}\), focusing early tuning on high-potential knobs while still supporting pruning and exploration.
\end{enumerate}

By integrating BO with RFECV and LRT in a feedback-driven cycle, \textsf{DOT} allocates tuning resources where they yield the greatest gains---minimizing overhead and maximizing performance across diverse DBMS workloads.
It thus addresses the following challenges:
1)\,\textbf{Identifying impactful knobs} — Done iteratively via RFECV using samples collected during tuning with minimal overhead.
2)\,\textbf{Workload dependency} — Adaptively adjusts based on workload specific feedback.
3)\,\textbf{Transferability vs.\ specificity} — Can use prior rankings but corrects them if needed via RFECV.
4)\,\textbf{Sampling overhead} — No up-front selection phase; tuning and selection are merged.
5)\,\textbf{Choosing how many knobs} — Controlled dynamically via LRT decisions and RFECV pruning.

\subsection{Bayesian Optimization via Gaussian Process}
\label{sec:bo}
\gls{bo} with GP\cite{bo_most} surrogate is our tuning algorithm of choice for expensive, noisy objectives~\cite{bo_ml1,bo_ml2,bo_drug1,Ottertune,ituned}.
In \textsc{DOT}, \gls{bo} is initialized with the current knob set $\mathcal{K}_{\mathrm{curr}}$, the number of optimization iterations $E$, and the set of prior samples $\mathcal{D}$ projected onto $\mathcal{K}_{\mathrm{curr}}$ as $\mathcal{D}_{\mathrm{proj}}$ (or to sample 10 random configurations if $\mathcal{D}$ is empty).
\gls{bo} first fits the GP model to $\mathcal{D}_{\mathrm{proj}}$, then for each of the $E$ iterations selects the next configuration via an acquisition function, evaluates it on the DBMS using adaptive benchmarking (Section~\ref{sec:budget_allocator}), and augments $\mathcal{D}_{\mathrm{proj}}$ with the resulting performance measurements.
After $E$ rounds, \gls{bo} returns the enriched set of configuration–performance pairs, often finding better configurations with far fewer evaluations than grid or random search, thereby reducing computational cost and enabling effective tuning under resource constraints.

\subsection{Recursive Feature Elimination with Cross-Validation}
\label{sec:rfecv}
RFECV is a widely adopted algorithm for feature selection in ML projects~\cite{RFECV_most}. 
\textsc{DOT} uses it to prune tuning knobs prior to optimization.
RFECV is initialized with the current knob set $\mathcal{K}_{\mathrm{curr}}$ and the projected sample set $\mathcal{D}_{\mathrm{proj}}$. 
An estimator---e.g., a random forest with CART as in~\cite{Hunter}---is trained on all candidate knobs; the least important knob(s) are removed and the model re‐evaluated via cross‐validation.
Unlike plain CART, which only ranks knobs and requires an arbitrary cutoff, RFECV leverages cross‐validation to automatically identify the optimal subset by iteratively eliminating features until the highest average CV score is achieved. To prevent over‐pruning and ensure sufficient search diversity, we enforce a minimum of 10 knobs in the pruned subset (avoid over-pruning to preserve BO diversity). By discarding knobs that minimally impact \gls{dbms} performance, RFECV reduces dimensionality and accelerates subsequent tuning. We adopt Scikit‐Learn’s RFECV implementation in \textsc{DOT} to focus the search on the most impactful knobs, thereby enabling faster optimizer convergence.

\subsection{Likelihood-Ratio Test for Set Expansion}
\label{sec:lrt-two-action}

In \textsc{DOT}, LRT\cite{vuong1989likelihood} is used to deterministically compare empirical success rates to select between two actions: \textbf{(0)} stay with current knobs or \textbf{(1)} expand the knob set. 
Each action \(i \in \{0,1\}\) tracks success and failure counts \((s_i, f_i)\), from which it computes a smoothed empirical success rate \(\hat{p}_i = s_i / (s_i+f_i)\). 
The decision is based on:
\[
 \Lambda \gets \ln\!\bigl(\hat p_1 / \hat p_0\bigr), \quad \text{action} = \mathds{1}[\Lambda > 0].
\]
This lightweight, adaptive rule requires only four scalars and no exploration schedule.

\vspace{0.5em}
\noindent\textbf{Reward function.}\ Given best performance \(b_{t-1}\), new value \(c_t\), and step count \(n_t\), define the per-call improvement as \(g_t = (c_t - b_{t-1}) / (b_{t-1} \cdot n_t)\). A binary reward is given by:
\[
r_t = \mathds{1}[g_t > \delta], \quad \text{e.g., } \delta = 0.001.
\]
We empirically set the threshold to 0.001 because, over 100 iterations, it corresponds to roughly a 10\% performance improvement, which is significant even with benchmarking noise.
The chosen action’s success/failure counts are updated accordingly. 
LRT selects to expand or to stay with the current knob set based on the feedback.

\begin{algorithm}[]
\caption{LRT for Knob Set Expansion}
\label{alg:lrt-stateful-flag}
\begin{algorithmic}[1]
  \Require
    Best past performance $b\in\mathbb{R}$;
    Best observed performance in the steps $y^*\in\mathbb{R}$;
    Number of iterations $E\in\mathbb{N}$
  \State \textbf{State:}
    Success/Failure counts $s_0,f_0,s_1,f_1$, all initialized to $1$;
    Flag $\mathit{init}\gets\text{True}$
  \State \textbf{Parameter:} reward threshold $\delta \gets 0.001$
  \Procedure{LRT\_Step}{$b,\,y^*,\,E$}
    \If{$\mathit{init}$}
      \State Draw $a_t \sim \mathrm{Bernoulli}(0.5)$
      \State $\mathit{init} \gets \text{False}$
      \State \Return $a_t$
    \EndIf
    \State $\hat p_0 \gets \dfrac{s_0}{s_0 + f_0},\quad \hat p_1 \gets \dfrac{s_1}{s_1 + f_1}$
    \State $\Lambda \gets \ln\!\bigl(\hat p_1 / \hat p_0\bigr)$
    \State $a_t \gets \mathds{1}[\Lambda > 0]$
    \State $g_t \gets \dfrac{y^* - b}{\,b\,E\,},\quad r_t \gets \mathds{1}[g_t > \delta]$
    \State $s_{a_t} \gets s_{a_t} + r_t,\quad f_{a_t} \gets f_{a_t} + (1 - r_t)$
    \State \Return $a_t$
  \EndProcedure
\end{algorithmic}
\end{algorithm}

The rationale behind the LRT is to drive exploration when the initial knob set yields poor performance improvement and to favor exploitation when the increase of performance is strong, avoiding unnecessary expansion of the search space, which would incur additional computational overhead.

\subsection{Adaptive Benchmarking}
\label{sec:budget_allocator}
Benchmarking can dominate tuning time
---e.g., 69.5\% for OLTP \textsc{Sysbench} with 20 knobs under BO, and 57\% for OLAP TPC-H scale-1 with 20 knobs under BO.
To address this, \textsf{DOT} employs an \emph{adaptive benchmarking} that early-terminates runs unlikely to surpass the best configuration, using partial results as proxies.
Below, we outline this approach for OLTP and OLAP workloads.

\begin{figure}[t!]
    \centering
    \includegraphics[
    width=0.9\linewidth,
    trim=0   1cm   0   0,
    clip
  ]{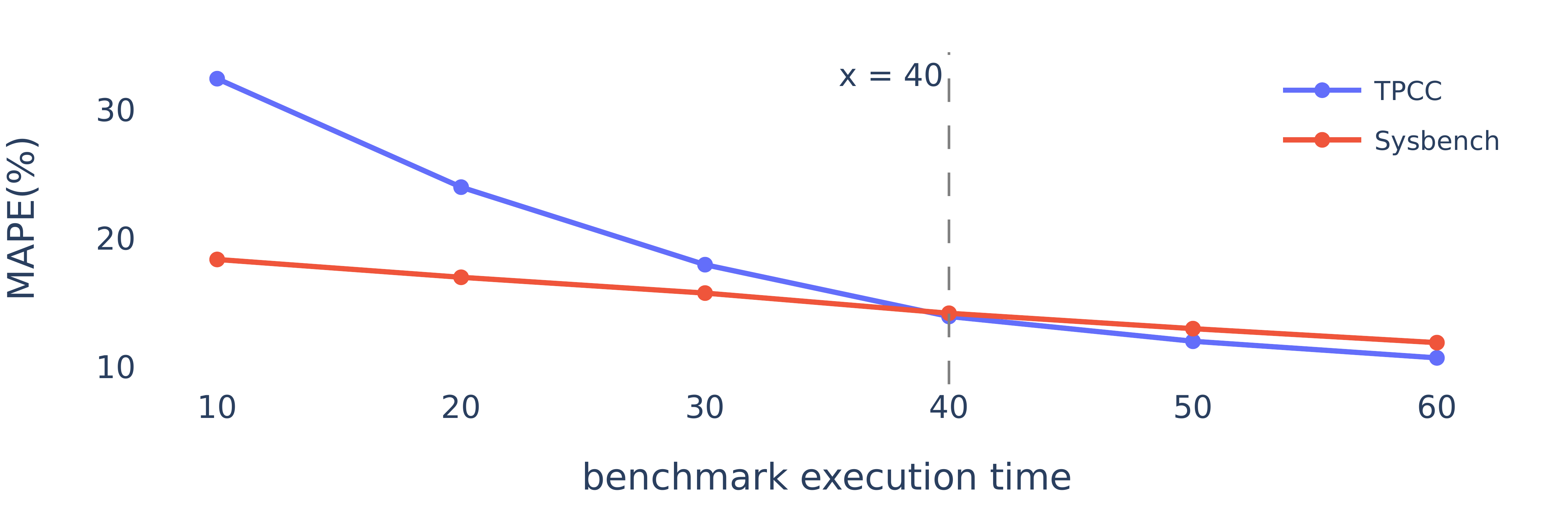}
    \caption{MAPE of Partial vs. Full Benchmark Results}
    \label{fig:MAPE}
\end{figure}

\textbf{OLTP Benchmarks.}
For workloads like \textsc{Sysbench} and \textsc{TPC-C}, performance is measured as steady-state throughput (TPS) after a warm-up phase. 
In our experiments, the canonical benchmark runs for $T_{\max}=90\,$s, using only the final $30\,$s to compute the \emph{ground-truth} TPS. 
To mitigate this, we run an early-cut window of $T_{\text{cut}}=40\,$s. 
If the average TPS at $T_{\text{cut}}$ already exceeds the current best, we let the benchmark continue to $T_{\max}$ for a precise measurement; otherwise, we abort and use the partial result as a (noisier) proxy 
. 
Figure~\ref{fig:MAPE} shows the \emph{Mean Absolute Percentage Error} (MAPE) between partial and full benchmark execution with more than $2,000$ runs under various configurations.
We observe that, at $T_{\text{cut}}=40\,$s, the 95\textsuperscript{th}-percentile MAPE is only $\SI{13.7}{\percent}$ for \textsc{TPC-C} and $\SI{14.2}{\percent}$ for \textsc{Sysbench}, with little benefit beyond this point.

\textbf{OLAP Benchmarks.} 
\setlength{\textfloatsep}{8pt}
\begin{figure}[h!]
    \centering
     \includegraphics[
    width=0.9\linewidth,
    trim=0   1cm   0   0,
    clip
  ]{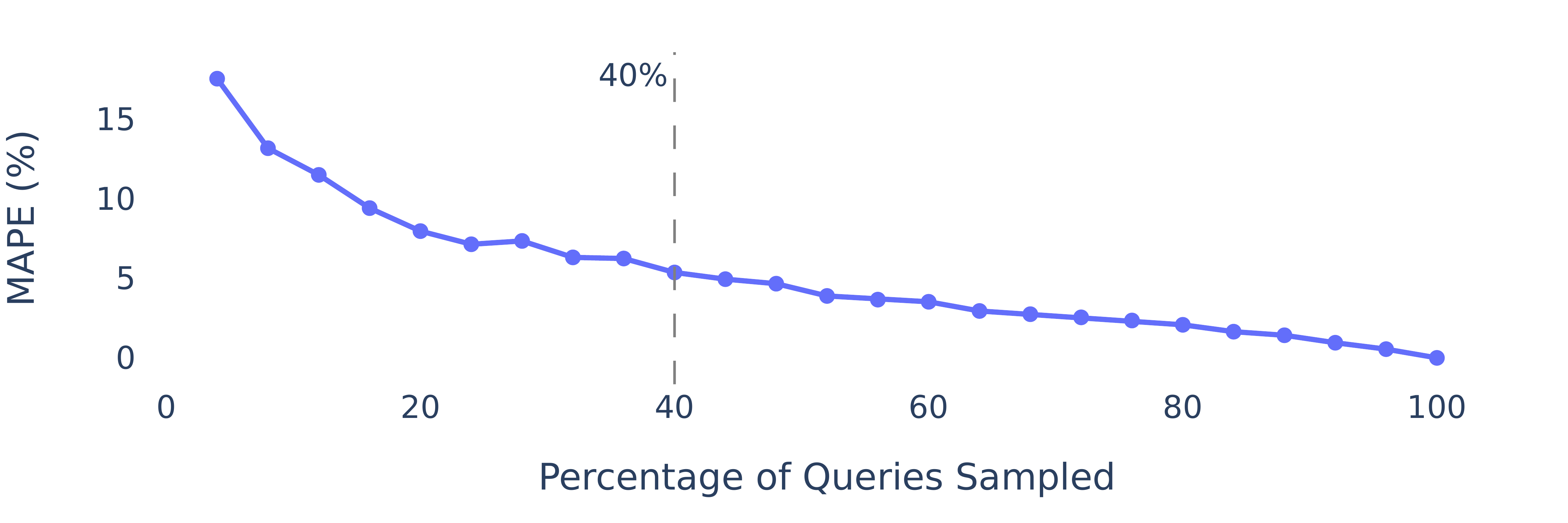}
    \caption{MAPE vs. Benchmark-Query Coverage}
    \label{fig:MAPE_olap}
\end{figure}
For workloads such as \textsc{TPC-H}, performance is the total execution time of a set of analytical queries, without warm-up. 
We randomly sample a subset of queries (40\% in our experiments) and track their execution time with the current configuration. 
If this partial workload completes faster than the best total time, we run the full workload to obtain the execution time; otherwise, we log the partial result as a proxy. 
Figure~\ref{fig:MAPE_olap} quantifies the error due to sampling over 1,000 logs under different configurations: sampling more queries reduces error, and 40\% sampling strikes a balance between accuracy and overhead reduction.


Note that while we empirically set our budget ($T_{\text{cut}} = 40$s and 40\% target query coverage) using historical data, it can be estimated from a few pilot measurements at minimal cost---Section~\ref{sec:ablation} shows that  performance is insensitive to the exact budget chosen.

\section{Experimental Results}\label{sec:baselines_and_experiments}
Besides \textsf{DOT}, we implemented two complementary baselines that \emph{dynamically} adjust the number of tuned knobs so as to lower the dimensionality of the search space and enhance the tuning speed.

\textbf{Incremental Knob Tuning.}
We follow \textsc{OtterTune}'s progressive expansion: BO starts on the top-4 knobs for 25 iterations, then appends the next two per epoch while reusing all observations; the process runs until the budget ends. This keeps early search low-dimensional but can stall if early rankings omit a key knob.

\textbf{Statistical Elimination (SE).}
SE proceeds inversely: start with all knobs, run $2N$ BO evaluations, then drop knobs that fail a per-knob Welch $t$-test ($\alpha{=}0.05$); BO continues only on the survivors (others are frozen at their best value). This upfront pruning reduces dimensionality and speeds subsequent search.

\subsection{Experimental Setup \& Results}
We evaluate \textsf{DOT}’s effectiveness and efficiency on the \textsc{Sysbench}, TPC-C, and TPC-H benchmarks, using the same workloads, DBMS, and hardware settings as in Section~\ref{sec:experiments_knob_selection}. Global space is also capped at the top 52 expert-selected parameters to simulate a practical tuning scenario. 
We compare \textsf{DOT} against:
\begin{itemize}[leftmargin=*]
  \item \textbf{BO}: Classical GP-BO as in iTuned~\cite{ituned} and \textsc{OtterTune}~\cite{Ottertune} incorporated with different knob sets: top 20 knobs ranked by experts (EXP), top 20 knobs ranked by CART model with 10 samples per knob (CART), and all the knobs in the scope (full);
  \item \textbf{Incremental Knob Tuning/ Statistical Elimination}: Our two dynamic‐search‐space variants;
  \item \textbf{\textsc{CDBTune}}: Author's public implementations of \textsc{CDBTune}~\cite{HustAIsGroup_CDBTune}, the implementation currently only supports OLTP;
  \item \textbf{DB-Bert}: Author's implementations of DB-Bert~\cite{itrummer_DB_BERT}, a LLM-based tuner, the implementation currently only supports OLAP;
  \item \textbf{LLM knowledge based}: An LLM‐based baseline adapted from \textsc{GPTuner}~\cite{GPTuner}, where we modified the prompt of \textsc{GPTuner}'s knob selection phase to let \textsc{ChatGPT} decide both which knobs and how many to tune.
\end{itemize}
Other methods, like \textsc{Hunter}, DDPG++(optimized version of \textsc{CDBTune} proposed in~\cite{DDPG++}), and UDO, are not included in our comparison: \textsc{Hunter}, DDPG++ lacks a public implementation, while UDO focuses on index tuning outside our scope.

Section~\ref{sec:experiments_knob_selection} showed that supplying a pre-ranked knob list greatly improves DBMS tuning by transferring prior knowledge, yet producing such rankings is costly, and they may be inaccurate. 
Consequently, we run each algorithm in two scenarios:
  \textbf{Expert ranking available.} We feed a trusted expert ranking to \textsf{DOT}, Incremental Tuning, BO limited to the Top\,20 expert‐ranked knobs, \textsc{CDBTune} restricted to those 20 knobs, and DB-BERT with its documentation‐derived priors;
  \textbf{Erroneous or missing ranking.} We provide a randomly permuted list (equivalent to no ranking) and evaluate \textsf{DOT}, Incremental Tuning, BO over the random list, BO over the full 52‐knob space, BO guided by LLM‐selected knobs, and BO+CART (counting CART’s ten samples per knob as a 14.4h overhead). 
  SE, which does not use rankings, is applied in both scenarios.

\subsection{Including Knob Importance Knowledge}
\begin{figure*}[t!]
    \centering
    \includegraphics[width=0.75
    \linewidth]{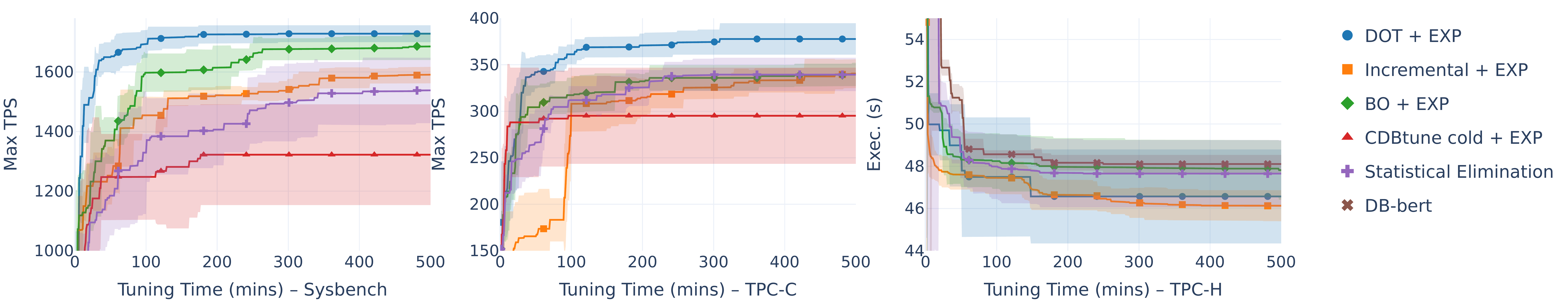}
    \caption{Performance vs.\ tuning‐time of different algorithms on \textsc{Sysbench}, TPC-C, TPC-H with knob importance knowledge}
    \label{fig:GENERAL_exp}
\end{figure*}

Figure~\ref{fig:GENERAL_exp} shows best–so–far performance versus tuning time for \textsc{Sysbench}, TPC–C and TPC–H when knob importance knowledge is available. 
On \textsc{Sysbench}, \textsf{DOT}+EXP reaches its optimal plateau faster than any other method and delivers throughput indistinguishable from BO+EXP. 
Other approaches—Incremental+EXP, \textsc{CDBTune}, and Statistical Elimination—settle at lower performance. 
\textsc{CDBTune}’s variability stems from its cold start; although a warm‐up phase can improve results, it adds significant overhead.

\setlength{\floatsep}{8pt}

On TPC-C, \textsf{DOT}+EXP not only converges quickly but also outperforms BO+EXP by a statistically significant margin, thanks to its iterative knob selection and occasional exploration of other knobs. 
All other methods remain clearly behind.
On TPC-H, confidence intervals overlap across methods, reflecting this workload’s relative insensitivity to tuning. 
Nevertheless, \textsf{DOT}+EXP still attains the lowest execution time, even if practical differences are small.
These results show that, when knob‐ranking priors are available, \textsf{DOT}+EXP consistently delivers the fastest convergence and either matches or exceeds the final performance of state‐of‐the‐art tuners.

\subsection{Excluding Knob Importance Knowledge}
\begin{figure*}[t!]
    \centering
    \includegraphics[width=0.75\linewidth]{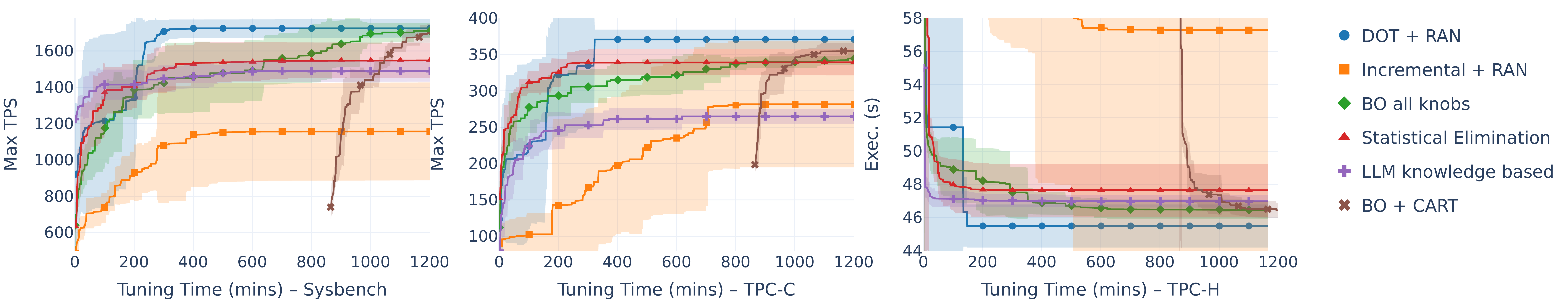}
    \caption{Performance vs.\ tuning‐time of different algorithms on \textsc{Sysbench}, TPC-C, TPC-H without knob importance knowledge}
    \label{fig:GENERAL_ran}
\end{figure*}

Figure~\ref{fig:GENERAL_ran} plots best–so–far performance when no knob‐importance prior is available.
\textsf{DOT} begins by exploring random knobs before focusing on the most important ones.
In \textsc{Sysbench} and TPC-C, \textsf{DOT+RAN} still converges the fastest. 
In \textsc{Sysbench}, it matches BO+All knobs and BO+CART and on TPC–C, it achieves a statistically superior throughput. 
For TPC-H, differences between methods are minor, though \textsf{DOT+RAN} attains the best average execution time. 
BO+CART climbs steeply once tuning starts—reflecting CART’s strong knob selection—but its costly pre-sampling adds significant overhead.
BO‐All knobs eventually plateaus alongside \textsf{DOT}, but only after a lengthy exhaustive search.
BO‐All knobs achieves similar performance as \textsf{DOT} yet takes a longer time; note that Incremental+EXP follows a similar staircase pattern with a lower ceiling as \textsf{DOT}. 
Both Incremental and \textsf{DOT} gradually incorporate new knobs, but \textsf{DOT} reaches higher scores more quickly.

We can also observe that \textsf{DOT+RAN} initially exhibits high variance, which steadily decreases as tuning progresses; as \textsf{DOT} explores more knobs, the set eventually converges to the most important parameters (Section~\ref{sec:behavior}).

\vspace*{-1em}\subsection{Comparison of All Methods \& Situations}


\begin{table*}[htbp]
  \caption{Performance, tuning time, and Friedman‐test ranks for different methods (statistically significant at $p <$ 5E-2)}
  \label{tab:merged_final_2}
  \centering
  \scriptsize
  \resizebox{\textwidth}{!}{%
    \begin{tabular}{@{}l|cc|cc|cc@{}}
      \toprule
      & \multicolumn{2}{c|}{\bfseries \textsc{Sysbench}}
      & \multicolumn{2}{c|}{\bfseries TPC-C}
      & \multicolumn{2}{c}{\bfseries TPC-H} \\
      \cmidrule(lr){2-3}\cmidrule(lr){4-5}\cmidrule(l){6-7}
      & \begin{tabular}{@{}c@{}}Rank (Max TPS)\\Test stat: 37.4\\p-value: 2.2E-05\end{tabular}
      & \begin{tabular}{@{}c@{}}Rank (tuning time [mins])\\Test stat: 32.2\\p-value: 1.9E-04\end{tabular}
      & \begin{tabular}{@{}c@{}}Rank (Max TPS)\\Test stat: 31.4\\p-value: 2.5e-04\end{tabular}
      & \begin{tabular}{@{}c@{}}Rank (tuning time [mins])\\Test stat: 37.9\\p-value: 1.7E-05\end{tabular}
      & \begin{tabular}{@{}c@{}}Rank (exec [s])\\Test stat: 18.382\\p-value: 3.1e-02\end{tabular}
      & \begin{tabular}{@{}c@{}}Rank (tuning time [mins])\\Test stat: 38.4\\p-value:  1.5E-05\end{tabular} \\
      \midrule
      \textsf{DOT}+EXP        &  \textcolor{red}{2.6 (1728.3 ± 28.3)}  &  \textcolor{red}{2.0 (57.7 ± 28.2)}  &  \textcolor{red}{1.2 (377.7 ± 16.9)}    &  2.4 (121.6 ± 72.1)  &  4.4 (46.6 ± 2.2)  &  3.4 (57.6 ± 54.3)  \\
      \textsf{DOT}+RAN        &      3.0 (1724.7 ± 50.6)               &  4.2 (171.1 ± 121.2)                   &  2.0 (370.8 ± 13.6)                    &  3.8 (187.3 ± 107.5)  &  \textcolor{red}{2.6 (45.5 ± 1.3)}  &  \textcolor{red}{2.0 (33.9 ± 63.9)}  \\
      LLM knowledge based     &      8.0 (1489.1 ± 57.5)               &  3.4 (129.3 ± 99.1)                     &  9.2 (264.9 ± 10.1)                  &  5.4 (288.0 ± 207.9)  &  6.6 (47.0 ± 0.4)  &  \textcolor{red}{2.0 (21.1 ± 10.0)}  \\
      BO+CART                 &      3.6 (1708.0 ± 29.1)               & 10.0 (1080.3 ± 32.6)                   &  5.0 (356.1 ± 9.4)                    &  9.8 (960.4 ± 57.6)  &  3.6 (46.4 ± 0.5)  & 10.0 (1027.4 ± 106.6) \\
      \textsc{CDBTune} cold+EXP        &      8.6 (1358.1 ± 152.9)              &  3.2 (218.6 ± 310.1)                   &  8.2 (295.2 ± 51.7)                    &  \textcolor{red}{1.2 (33.5 ± 40.9)}  &      -                 &             -          \\
      DB-bert                 &      -                                  &      -                                 &              -                       &               -       &  8.6 (48.1 ± 0.4)  &  5.0 (114.0 ± 58.3)  \\
      Statistical Elimination &      7.2 (1548.4 ± 97.3)                &  4.8 (231.8 ± 103.5)                   &  6.4 (339.4 ± 18.2)                  &  3.2 (155.7 ± 50.0)  &  6.0 (47.6 ± 1.6)  &  4.2 (79.8 ± 52.5)   \\
      BO All Knobs                 &      3.0 (1720.0 ± 33.0)                &  8.8 (732.6 ± 266.7)                  &  6.0 (348.4 ± 10.1)                    &  7.8 (560.4 ± 327.7)  &  4.4 (46.5 ± 0.6)  &  7.8 (334.1 ± 175.5) \\
      BO+EXP                  &      2.8 (1725.2 ± 31.1)                &  5.0 (217.8 ± 42.7)                   &  4.8 (354.2 ± 14.8)                    &  6.2 (411.9 ± 264.9)  &  7.4 (47.8 ± 1.4)  &  4.6 (73.0 ± 53.2)   \\
      Incremental+RAN         &      9.6 (1157.2 ± 269.5)               &  7.0 (353.9 ± 86.0)                   &  7.4 (281.6 ± 86.8)                   &  8.0 (625.1 ± 113.6)  &  7.2 (57.3 ± 14.5) &   8.6 (432.2 ± 74.0) \\
      Incremental+EXP         &      6.6 (1620.0 ± 45.7)                &  6.6 (260.2 ± 82.7)                   &  4.8 (352.5 ± 18.5)                   &  7.2 (434.4 ± 149.0)  &  4.2 (46.1 ± 0.7)  &   7.4 (236.5 ± 40.5) \\
      \bottomrule
    \end{tabular}%
  }
\end{table*}
  
Table~\ref{tab:merged_final_2} presents Friedman‐test ranks for both performance (throughput or execution time) and tuning time across \textsc{Sysbench}, TPC–C and TPC–H. 
In every case, \textsf{DOT} variants occupy the top ranks and dramatically reduce tuning effort compared to BO+EXP, as it has a generally good performance across workloads, while other baselines lag in quality, speed, or both. 

On \textsc{Sysbench}, \textsf{DOT}+EXP achieves a performance rank of 2.6 and a tuning‐time rank of 2.0, delivering 1\,728.3 ± 28.3 TPS in 57.7 ± 28.2 min versus BO+EXP’s 1\,725.2 ± 31.1 TPS in 217.8 ± 42.7 min—cutting tuning time by 73.6\%. 
Even without priors, \textsf{DOT}+RAN (rank 3.0/4.2) matches final throughput (1\,724.7 ± 50.6 TPS) in 171.1 ± 121.2 min, outperforming other baselines and matches BO+EXP. 

On TPC-C, \textsf{DOT}+EXP leads with ranks 1.2 (performance) and 2.6 (time), achieving 377.7 ± 16.9 TPS in 121.6 ± 72.1 min—a 70.5\% reduction in tuning time compared to BO+EXP’s 411.9 ± 264.9 min. 
DOT+RAN (2.0/3.8) converges to 370.8 ± 13.6 TPS in 187.3 ± 107.5 min, while BO+CART, \textsc{CDBTune}, and DB‐bert remain behind.

On TPC-H, \textsf{DOT}+RAN ranks 2.6/2.0 by reaching 45.5 ± 1.3 s in 33.9 ± 63.9 min, surprisingly edging \textsf{DOT}+EXP (4.4/3.4) at 46.6 ± 2.2 s in 57.6 ± 54.3 min due to TPC-H's low tuning sensitivity for different knobs. 
Both outperform BO+EXP (7.4/4.6) by shaving 21.1\% (\textsf{DOT}+EXP) and 53.6\% (\textsf{DOT}+RAN) off its 
tuning time.

State-of-the-art methods, like \textsc{CDBTune} and DB-BERT perform poorly here because we deliberately simulate a low-knowledge scenario. 
\textsc{CDBTune} requires an extended warm-up phase to gather data, and DB-BERT needs more complete documentation and carefully tuned hyperparameters. 
Under our “cold start” conditions, neither can exploit such prior information, so their performance lags. 
\textbf{LLM-based methods} face similar limitations. Approaches such as GPTuner-style prompt ranking rely on access to comprehensive DBMS manuals and often require multiple exchanges with the LLM to refine the ranking. 
While these techniques can reduce sampling overhead when full documentation is available, in our cold-start setting, they cannot leverage sufficient prior knowledge, leading to weaker performance.
 
Across all benchmarks, \textsf{DOT}+EXP and \textsf{DOT}+RAN not only secure the best or near‐best ranks but also slash tuning time by up to 73.6\% (on \textsc{Sysbench})
relative to BO+EXP, underscoring their efficiency and robustness compared to every other method. 

These results underscore two key strengths of \textsf{DOT}.
First, even without priors, \textsf{DOT}+RAN can match or exceed BO+EXP by dynamically discovering and exploiting the most influential knobs based on the current workload. 
Second, when knob‐importance priors are available, \textsf{DOT}+EXP leverages them to converge even faster—on \textsc{Sysbench} and \textsc{TPC-C} with minimal overhead compared to \textsf{DOT}+RAN. 
In both cases, \textsf{DOT} delivers high‐quality tuning with little extra cost, making it a practical choice whether or not prior knowledge is at hand.
\setlength{\textfloatsep}{8pt}
\begin{figure}[]
   \centering
  \includegraphics[width=0.85\linewidth]{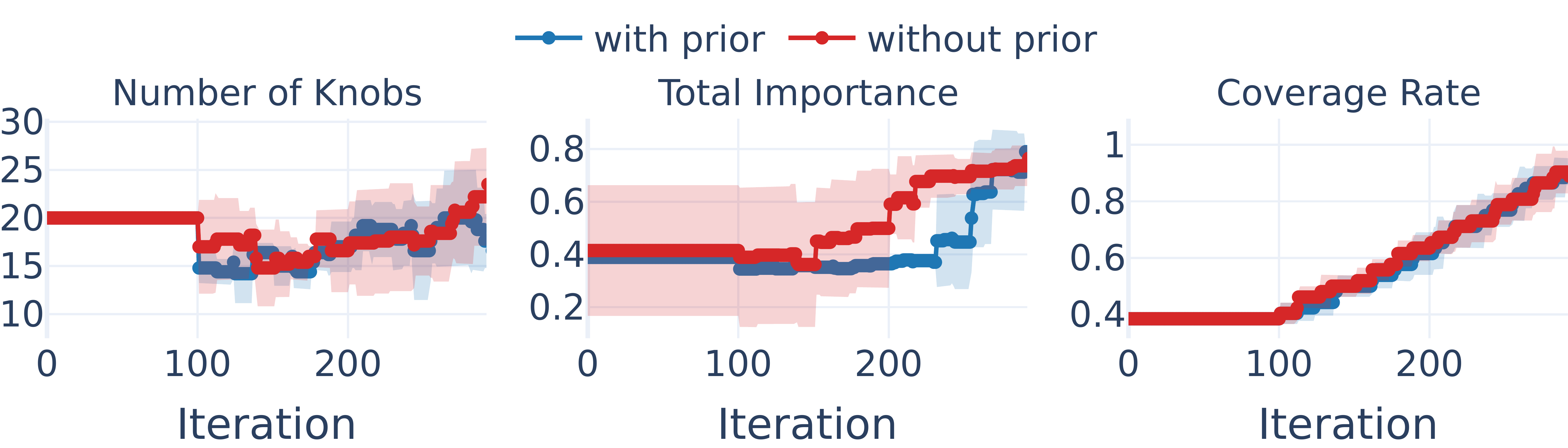}
    \caption{DOT behavior on \textsc{Sysbench} workload with and without knob importance prior}\label{fig:behavior}
\end{figure}

\subsection{DOT Behavior Analysis}
\label{sec:behavior}
Figure~\ref{fig:behavior} shows \textsf{DOT}’s tuning behavior on a \textsc{Sysbench} workload, comparing runs with an expert‐ordered knob‐importance prior versus a randomized start (no prior). 

\textbf{Dynamic Tuned Knobs Count.}
As iterations proceed, \textsf{DOT} applies RFECV to prune low‐value knobs and uses LRT to decide when to expand its search. 
Without a prior, more knobs are eliminated in early phases, since the random initial set has a lower average importance. 
In both setups, \textsf{DOT} settles on tuning roughly 20 knobs to avoid excessive overhead.

\textbf{Total Importance of Tuned Knobs.}
With a prior, the cumulative importance of the active knobs remains nearly constant at first—high‐impact variables cycle in and out but stay within the tuning pool---and then increases in a step-wise fashion.
Without a prior, the initial knob set exhibits higher variance in importance, and its cumulative score rises quickly, eventually converging with the prior case. 
This occurs because, freed from expert bias, the no‐prior run discovers impactful knobs discovered by CART that the expert had down‐ranked. 
This highlights the trade‐off between transferability and specificity. 
Incorporating a prior on knob importance can come at the expense of specificity, whereas a purely random search may more quickly home in on the most critical knobs. 
However, even though the “no-prior” approach can yield higher importance scores, it does not necessarily translate into better final performance, since CART-derived importances are not guaranteed to reflect the true ground truth.
By the end, both algorithms retain about 20 knobs whose total importance exceeds 0.8 (out of 1).

\textbf{Coverage Rate.}
Under both scenarios, \textsf{DOT} covers the majority of the knob space.
The no‐prior case achieves higher coverage initially, 
demonstrating that the LRT favors exploration in the absence of informative priors while concentrating exploitation when such information is available. 
And once exploitation yields diminishing returns, \textsf{DOT} transitions to exploration for further fine-tuning. 


In summary, \textsf{DOT} meets its goals—balancing exploration and exploitation, keeping the search space compact, and minimizing overhead—while preserving key knobs.

\subsection{Periodic Workload}
\begin{figure}[t!]
  \centering
  \includegraphics[
    width=0.78\linewidth,
    trim={0   0.5cm   0   0.5cm},
    clip
  ]{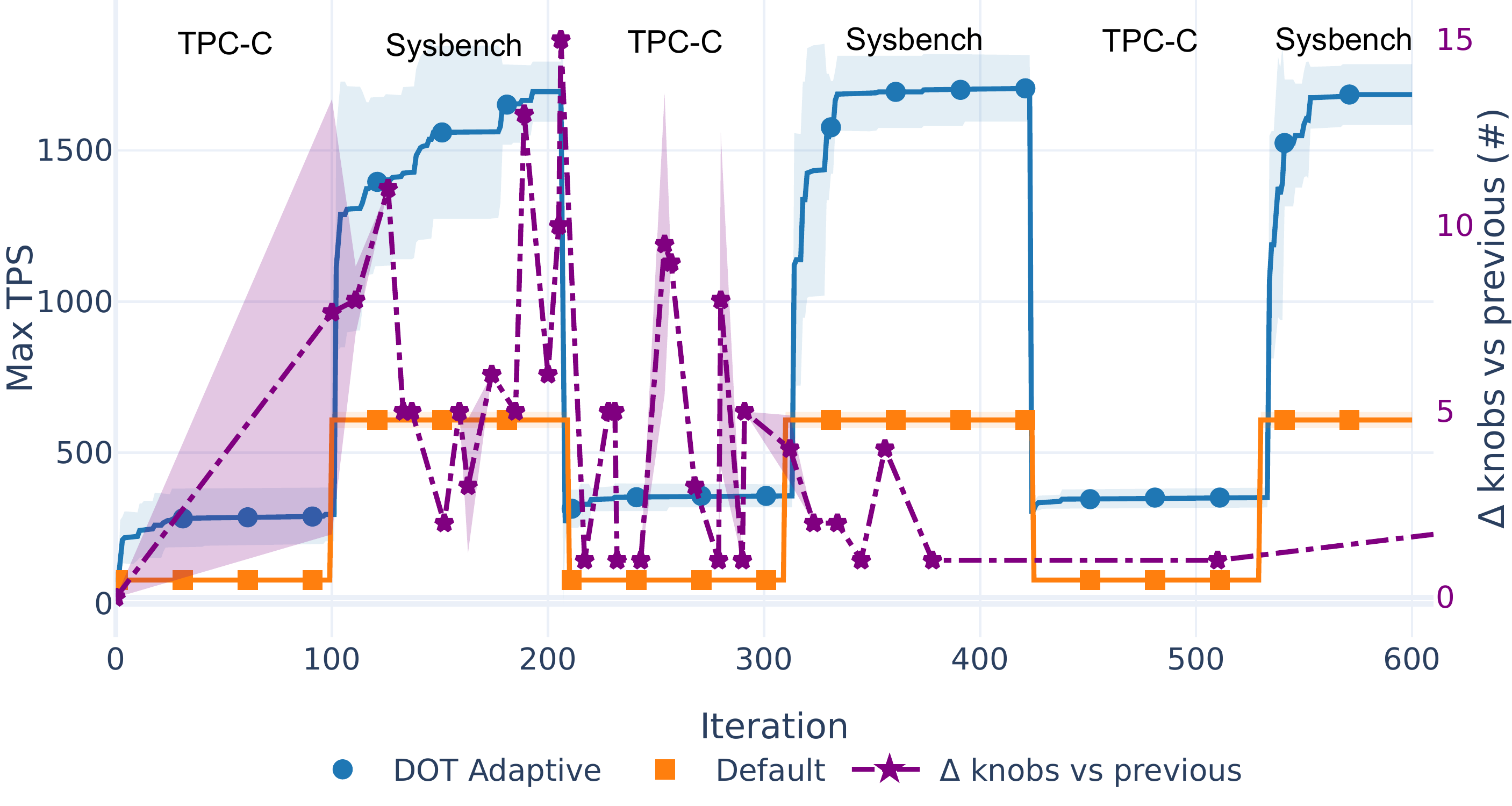}
  \caption{Tuning Performance of \textsf{DOT} on periodic workloads composed of TPC-C and Sysbench}
  \label{fig:periodic}
\end{figure}


\begin{figure}[t!]
  \centering
  \includegraphics[
    width=0.7\linewidth,
    trim=0.15cm   0.6cm   0.7cm   1cm,
    clip
  ]{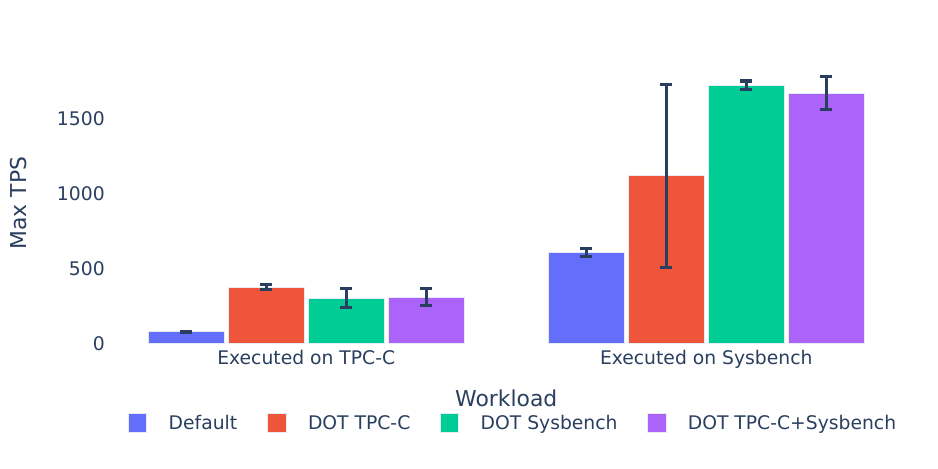}
  \caption{Max TPS Across Configurations and Workloads}
  \label{fig:periodic_2}
\end{figure}

In addition to static workloads, we also evaluate \textsc{DOT} under a periodic workload mixture, where execution alternates between TPC-C and Sysbench following the protocol of \cite{periodic}. Figure~\ref{fig:periodic} illustrates this dynamic scenario. The \textsc{DOT Adaptive} curve shows the behavior of the tuning procedure, which adapts dynamically to workload changes and progressively accelerates convergence to high-performance configurations after each shift. The dotted curve represents the number of knobs modified between consecutive iterations. It initially shows a high volume of changes to adapt to both workloads, which eventually stabilizes, indicating that \textsc{DOT} adapts effectively while converging toward a stable, efficient knob set.

Figure~\ref{fig:periodic_2} shows the best configuration found by \textsc{DOT} when trained on the periodic workload, denoted as \textsc{DOT TPC-C+Sysbench}. Its throughput is lower than that of workload-specific configurations, as it must jointly optimize across multiple workloads without a mechanism to distinguish them. Nevertheless, \textsc{DOT TPC-C+Sysbench} consistently outperforms both the default configuration and mismatched static baselines, and achieves performance that is statistically indistinguishable from \textsc{DOT SYSBENCH}. This demonstrates that \textsc{DOT} can be effectively trained on workload mixtures, delivering robust performance.

\subsection{Ablation Study}
\label{sec:ablation}


\begin{figure}[t!]
  \centering
  \includegraphics[
    width=0.7\linewidth,
    trim=0   0cm   0   0,
    clip
  ]{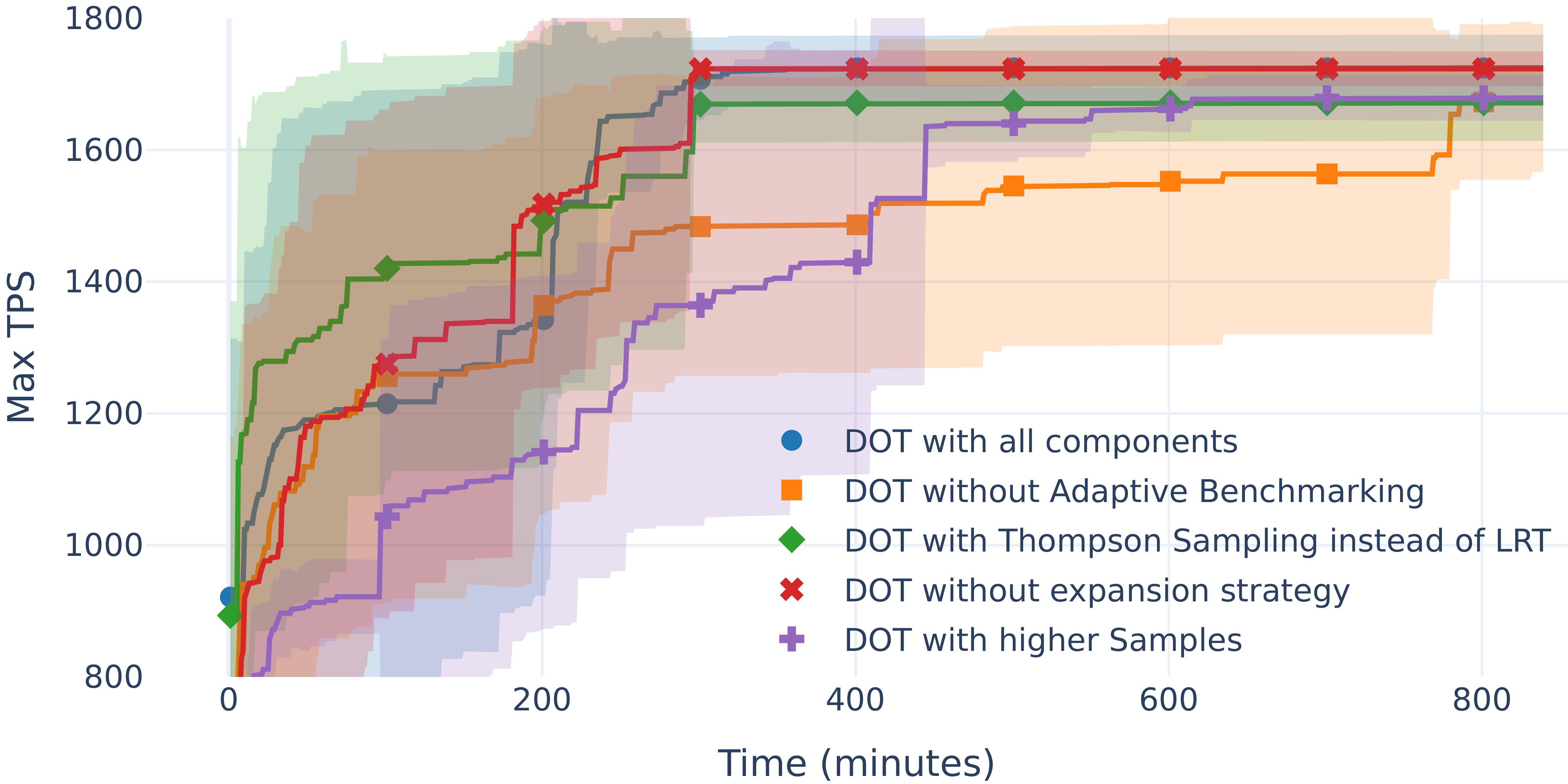}
  \caption{Tuning Performance of \textsf{DOT} on \textsc{Sysbench} for various configurations without knob importance knowledge}
  \label{fig:ablation}
\end{figure}

\begin{figure}[t!]
  \centering
  \includegraphics[
    width=0.76\linewidth,
    trim=0   0cm   0   0,
    clip
  ]{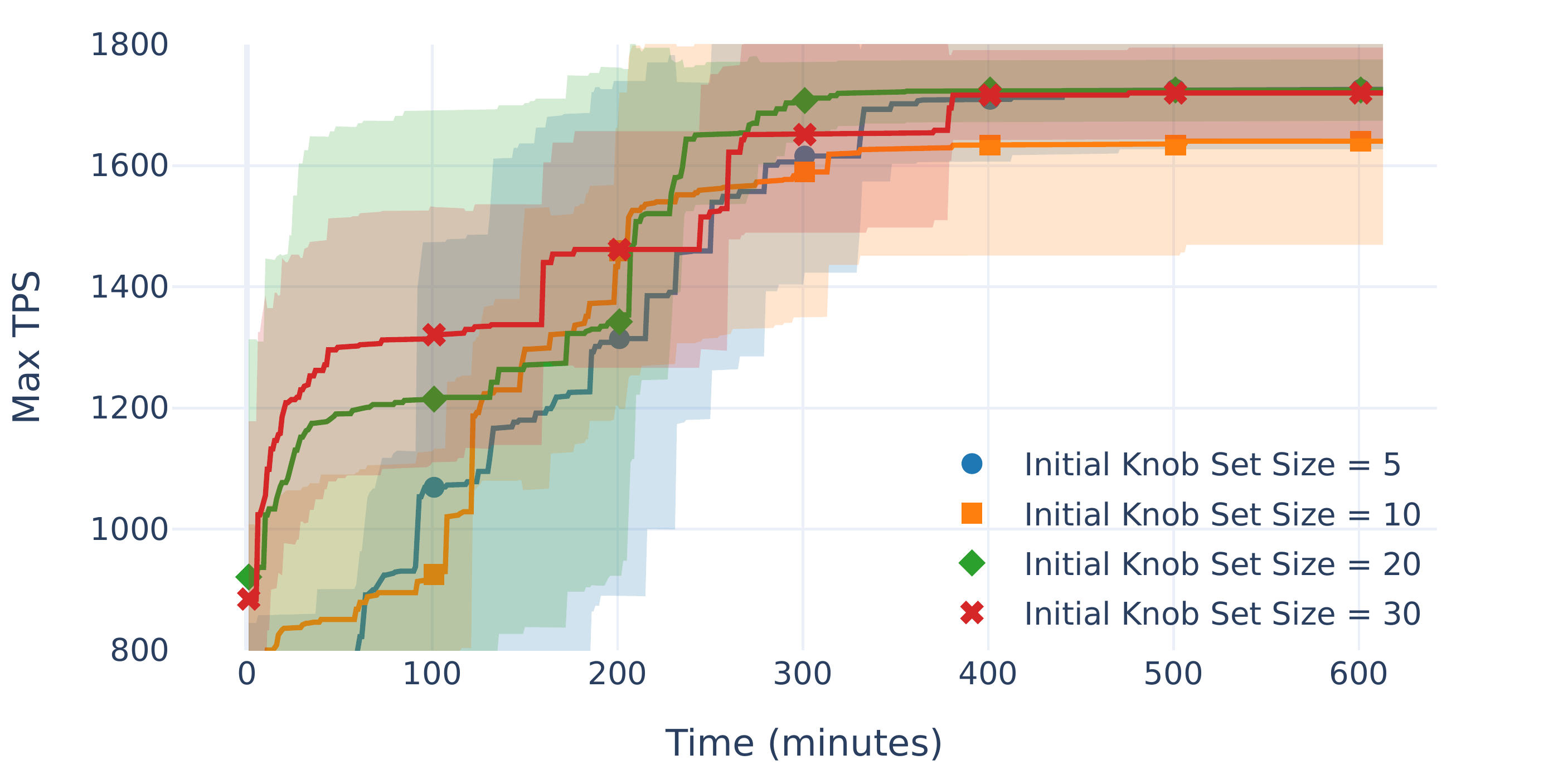}
  \caption{Tuning Performance of \textsf{DOT} on \textsc{Sysbench} for various $k_0$ without knob importance knowledge}
  \label{fig:ablation_k}
\end{figure}

\begin{figure}[t!]
  \centering
  \includegraphics[
    width=0.7\linewidth,
    trim=0   0cm   0   0,
    clip
  ]{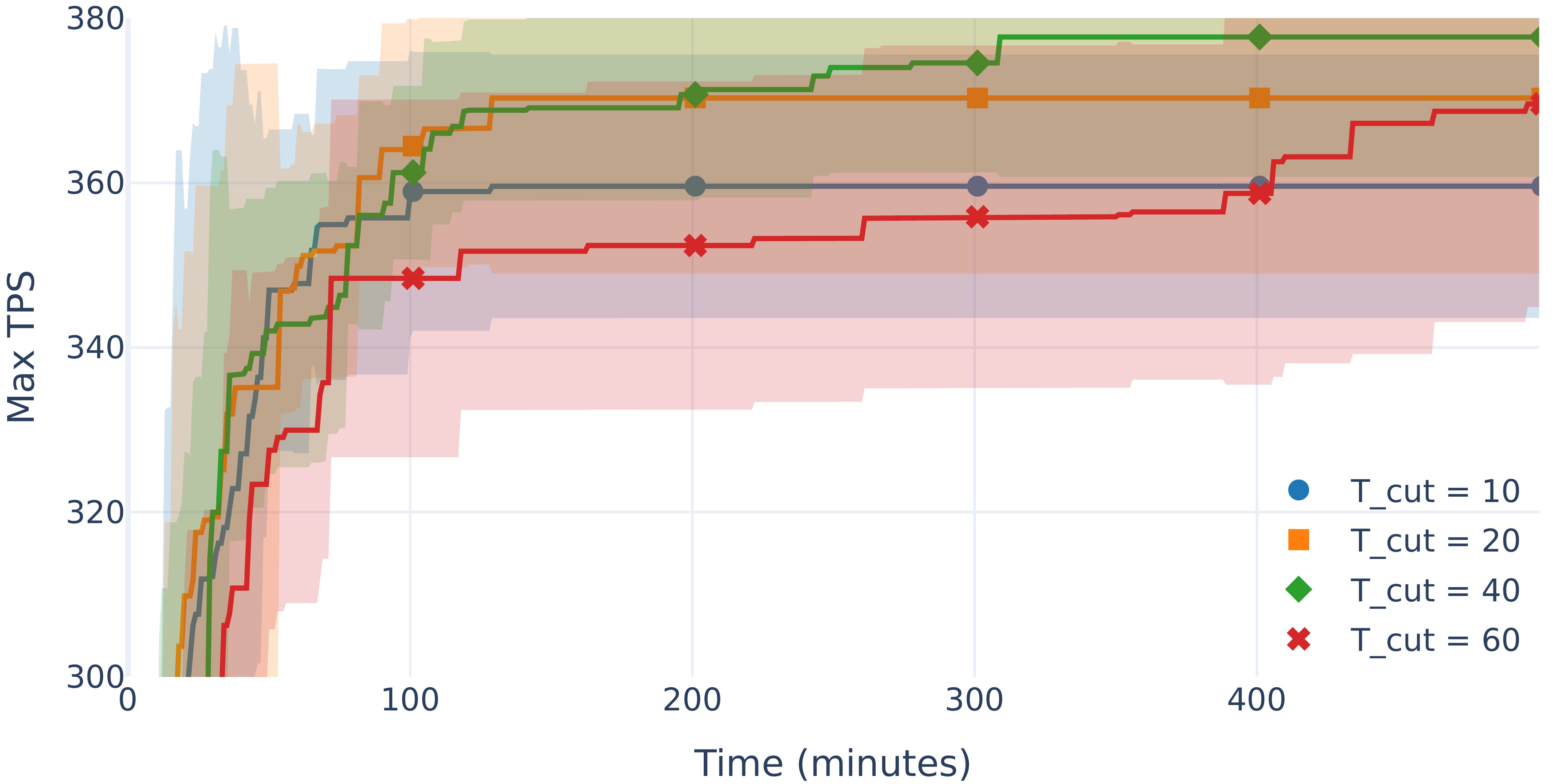}
  \caption{Tuning Performance of \textsf{DOT} on \textsc{TPC-C} for various $T_{\text{cut}}$ budget with knob importance knowledge}
  \label{fig:ablation2}
\end{figure}

In this section, we evaluate the performance of \textsf{DOT} on \textsc{Sysbench} by ablating its key components.
Figure~\ref{fig:ablation} demonstrates \textsc{DOT} performance under different components for Sysbench workload without any prior knob‐importance knowledge:
\begin{enumerate}[leftmargin=*]
\item \textbf{\textsf{DOT} without adaptive benchmarking:} Removing adaptive benchmarking slows convergence: the best‐so‐far curve climbs more gradually and only reaches its plateau later, although the final throughput remains unchanged. This demonstrates \textsf{DOT}’s resilience to noisy performance estimates.
\item \textbf{\textsf{DOT} with Thompson sampling / without expansion strategy:}  Replacing our LRT‐based knob set expansion strategy with Binary Thompson Sampling~\cite{ts}, or eliminating expansion strategy (i.e., exploring knobs purely incrementally), yields virtually statistically equivalent results according t-test. This indicates that \textsf{DOT}’s overall framework is robust to the choice of knob‐selection heuristic.
Naturally, purely incremental exploration performs well when no initial knob ranking is available, but LRT helps prevent over-exploration when knob‐importance priors exist (Section~\ref{sec:behavior}). 
Thus, LRT serves as a generally applicable method, though other strategies can be plugged into the \textsf{DOT} framework to suit specific tuning scenarios.
\item \textbf{\textsf{DOT} with higher samples:}  Doubling BO epochs per knob—from 5 to 10—to feed RFECV more samples for pruning delivers no performance gain but delays convergence to the final performance plateau. Therefore, 5 epochs per knob offer a practical trade-off between RFECV accuracy and run-time efficiency.
\end{enumerate}

Similarly, Figure~\ref{fig:ablation_k} shows the performance of \textsf{DOT} under different initial knob sizes ($k_0$) on a Sysbench workload without prior knowledge. Larger $k_0$ values (20, 30) yield stronger initial performance, whereas smaller values (5, 10) start lower but still converge at roughly the same rate. This suggests that \textsf{DOT} remains insensitive to the initial number of knobs, though selecting an appropriate value (e.g., 20) can provide an early advantage.
 
Additionally, we evaluated \textsf{DOT} on TPC-C (which had the highest MAPE in Section~\ref{sec:budget_allocator}) using prior knob‐importance knowledge and varying the cutoff budget \(T_{\text{cut}}\). 
Figure~\ref{fig:ablation2} shows that larger \(T_{\text{cut}}\) values slow convergence—since each benchmark run takes longer (e.g., \(T_{\text{cut}}=60\) converges much later than \(T_{\text{cut}}=10\)). 
However, even a tiny budget (\(T_{\text{cut}}=10\)) delivers final tuning quality nearly indistinguishable from that of high budgets for this \textbf{most noisy} benchmark. 
This insensitivity arises because our allocator 
concentrates precise measurements on promising configurations and uses coarser estimates elsewhere, letting BO exploit the best regions even without a finely tuned overall budget. 
Although approximate benchmarking can weaken the surrogate and require additional BO iterations---so a carefully chosen budget may yield faster convergence and better results---the end‐to‐end tuning time and ultimate performance remain rather robust to different \(T_{\text{cut}}\). To summarize, these components collectively provide a robust, general‐purpose solution for DBMS tuning.

\subsection{System Overhead}
\begin{figure}[t!]
    \centering
    \includegraphics[width=0.8\linewidth]{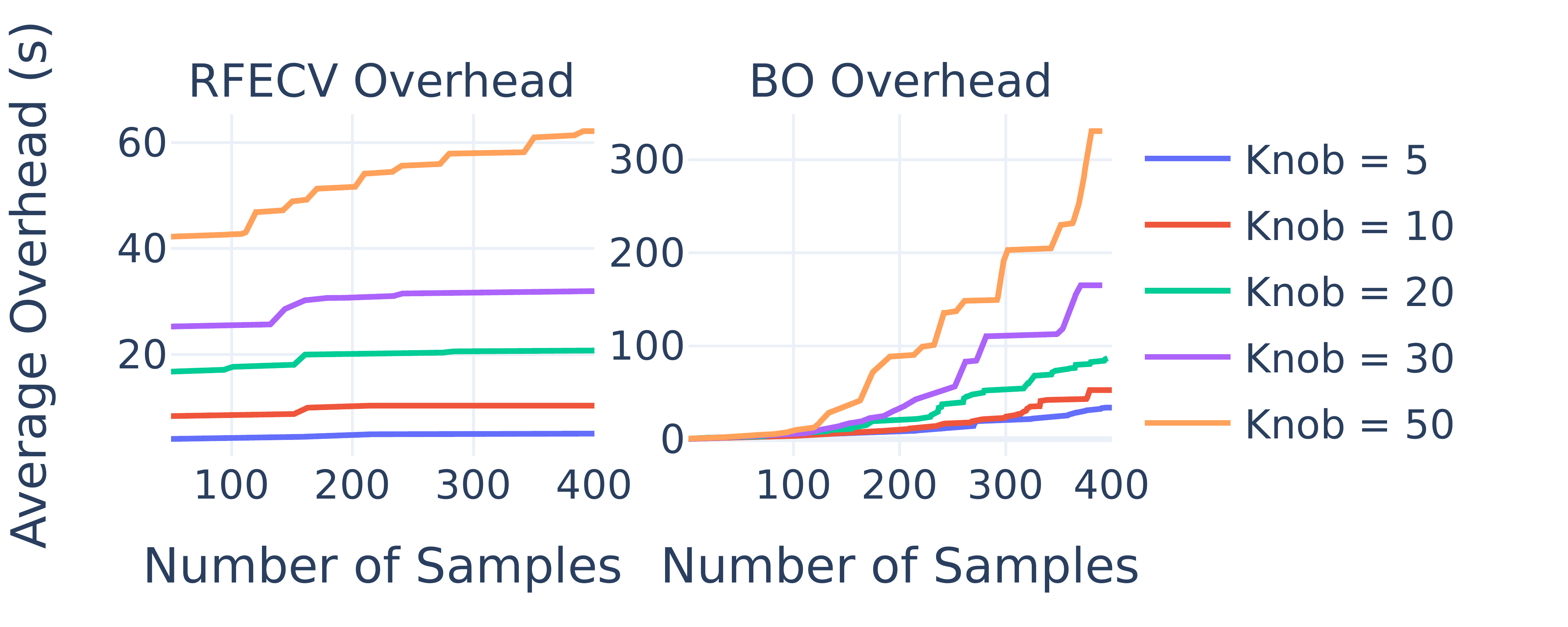}
    \caption{Bayesian optimization \& \gls{rfecv} overhead}
    \label{fig:bo_overhead}
\end{figure}

The BO shows rapidly growing cost when the number of samples increases: once the knob count exceeds about 20, the GP‐fit and acquisition step dominates, climbing from seconds to several minutes as samples accumulate (Figure~\ref{fig:bo_overhead}). 
By contrast, RFECV—used only to prune knobs before the main search—adds a small, knob‐dependent fixed tax that flattens after the first hundred samples and remains almost insensitive to further data, resulting in relatively low overhead.
Across all tested scenarios, benchmarking dominates the wall‐clock budget (56\%), BO bookkeeping follows at around 27\%, RFECV pruning contributes only about 0.4\%, and the remaining other tasks (logging, configuration deployment, DBMS restarts, LRT training and inference) consume roughly 16\%. Thanks to LRT’s simple structure, its training and inference cost is negligible (<0.01\%). This lightweight overhead of RFECV and LRT makes \textsf{DOT} more adaptable for wider adoption.
\setlength{\textfloatsep}{8pt}

\section{Conclusion \& Future Works}
We introduced \textsf{DOT}, a database autotuning framework that needs neither upfront training nor knob-importance priors, yet can exploit them when available.
Combining Bayesian optimization for strong proposals, RFECV with online sampling to prune low-impact knobs, and LRT to balance exploration and exploitation, \textsf{DOT} maintains a compact search space and avoids the curse of dimensionality.
Across diverse workloads, it matches or surpasses state-of-the-art tuners while cutting tuning time by orders of magnitude.
Even without reliable importance scores, \textsf{DOT} consistently identifies and focuses on the most impactful knobs.

Future directions include: (i) incorporating workload characterization~\cite{workload_detector1, workload_detector2, Ottertune} and other optimzaition objectives to better adapt to dynamic environments; (ii) exploring additional component optimizations or alternative heuristics to extend applicability; (iii) integrating more advanced hyperparameter optimization engines~\cite{smac_dbms}; and (iv) validating DOT’s scalability on real-world workloads and across multiple DBMSs. 
Overall, \textsf{DOT} demonstrates that adaptive dimensionality reduction combined with principled exploration can make database autotuning both efficient and practical.

\clearpage
\bibliographystyle{ACM-Reference-Format}
\bibliography{sample}

\end{document}